\documentclass[acmsmall,screen]{acmart}
\usepackage{natbib}
\usepackage[]{mdframed}
\AtBeginDocument{%
  }

\setcopyright{cc}
\setcctype{by}
\acmJournal{PACMHCI}
\acmYear{2026} \acmVolume{10} \acmNumber{6} \acmArticle{CSCW167}
\acmMonth{10} \acmDOI{10.1145/3817015}

\def\citepos#1{\citeauthor{#1}'s \citep{#1}}

\begin{document}

\title{The Relational Origins of Rules in Online Communities
}%

\author{Charles Kiene}
\orcid{0000-0002-7327-8391}
\email{ckiene@uw.edu}
\affiliation{%
  \institution{University of Washington}
  \city{Seattle, WA}
  \country{USA}}

\author{Sohyeon Hwang}
\orcid{0000-0001-8415-7395}
\email{sohyeon@princeton.edu}
\affiliation{%
  \institution{Princeton University}
  \city{Princeton, NJ}
  \country{USA}}

\author{Nathan TeBlunthuis}
\orcid{0000-0002-3333-5013}
\email{nathante@uw.edu}
\affiliation{%
  \institution{University of Texas, Austin}
  \city{Austin, TX}
  \country{USA}}

\author{Carl Colglazier}
\orcid{0000-0002-9068-8827}
\email{carlcolglazier@u.northwestern.edu}
\affiliation{%
  \institution{Northwestern University}
  \city{Evanston, IL}
  \country{USA}}

\author{Aaron Shaw}
\authornote{Authors contributed equally and order in author list was randomized.}
\email{aaronshaw@northwestern.edu}
\orcid{0000-0003-4330-957X}
\affiliation{%
  \institution{Northwestern University}
  \city{Evanston, IL}
  \country{USA}}

\author{Benjamin Mako Hill}
\authornotemark[1]
\email{makohill@uw.edu}
\orcid{0000-0001-8588-7429}
\affiliation{%
  \institution{University of Washington}
  \city{Seattle, WA}
  \country{USA}}

\renewcommand{\shortauthors}{}

\begin{abstract}
Where do rules come from in online communities? 
This study investigates how and why online communities adopt and change their rules. We conducted a grounded theory-based analysis of 40 in-depth interviews with community leaders from subreddits, Fandom wikis, and Fediverse servers, and identified seven processes involved in the adoption of online community rules. Our findings reveal that, beyond  \textit{operational} reasons like regulating behavior and solving problems, rules are also adopted and changed for \textit{relational} reasons, such as signaling or reinforcing community legitimacy and identity to other communities. While rule change was often prompted by challenges during community growth or decline, change also depended on volunteer leaders' work capacity, the presence of member feedback mechanisms, and relational dynamics between leaders and members. Our findings extend prior theories from social computing and organizational research, illustrating how institutionalist and ecological explanations of the relational origins of rules complement operational accounts. Finally, we build on these explanations to offer a set of design propositions that reflect the relational aspects of rules and rulemaking across communities' lifecycles.
\end{abstract}

\maketitle

\section{Introduction}
Rules serve many purposes in online communities. Rules codify social norms \citep{chandrasekharan_internetshidden_2018,fiesler_redditrules_2018}, signal appropriate behavior \citep{matias_preventingharassment_2019}, and become sites of contesting values \citep{gillespie_custodiansinternet_2018, stockinger_navigatinggray_2023,jhaver_doestransparency_2019,crawford_whatflag_2016}. Prior studies in social computing characterize extant rules \citep[e.g.,][]{fiesler_redditrules_2018,nicholson_mastodonrules_2023,hwang_rulesrulemaking_2022, keegan_evolutionconsequences_2017,reddy_evolutionrules_2023} as well as how they are enforced \citep[e.g.,][]{dosono_moderationpractices_2019,wohn_volunteermoderators_2019,cullen_practicingmoderation_2022,matias_preventingharassment_2019}, including the impacts of automated moderation and enforcement tools \citep[e.g.,][]{chandrasekharan_crossmodcrosscommunity_2019,teblunthuis_effectsalgorithmic_2021,halfaker_oreslowering_2020}. %
However, most studies take for granted that rules must also be \emph{designed} and \emph{adopted}. %
How and why do rules come to be? What challenges do leaders face in creating and changing them?

Prior work in social computing offers two explanations of the origins of community rules. The first emphasizes decision-making by platform owners or designers who assume rulemaking authority and seek to engineer or govern social interaction in particular ways \citep[e.g.,][]{kraut_buildingsuccessful_2012, gillespie_custodiansinternet_2018}. The second emphasizes localized rule development in response to specific challenges, conflicts, or constraints \citep[e.g.,][]{frey_thisplace_2019, seering_whomoderates_2022}. While these approaches differ in their objects of study, both treat rules as primarily \textit{operational}---i.e., created to solve practical problems and structure day-to-day activity within communities.
Previous empirical work has shown that many online communities have similar rules \citep{reddy_evolutionrules_2023, fiesler_redditrules_2018, nicholson_mastodonrules_2023, hwang_rulesrulemaking_2022,chandrasekharan_internetshidden_2018}.
Both approaches would thus assume that similar rules arise from similar problems. However, other prior work casts doubt on this and suggests that rules originate for non-operational reasons as well \citep[e.g.,][]{hwang_adoptingthirdparty_2024, hwang_rulesrulemaking_2022, teblunthuis_nocommunity_2022}.

Although there is substantial theoretical, descriptive, and quantitative research into online community rules, we lack a grounded understanding of how rule similarity arises from the processes by which communities adopt or change their rules.
To understand the origins of rules in online communities, we conducted 40 in-depth, open-ended interviews with leaders of communities from three different sociotechnical systems: subreddits on Reddit, wikis on Fandom, and servers in the Fediverse. In order to investigate the sources of rule similarity, we recruited interviewees from clusters of communities with similar rules in each setting. We report findings based on a comparative, grounded theory analysis \citep{charmaz_constructinggrounded_2015}, identifying seven distinct processes involved in rulemaking in online communities. 

Across these processes, we find that a community's rules and rulemaking are not only operational, governing their internal affairs, but are also importantly \textit{relational}, navigating their external constituencies and constraints.\footnote{Our use of the term ``relational'' is distinct from the way it is used within sociology to describe, for example, theories of relationalism versus substantivalism.} To community leaders, rules help draw distinctions between communities, manage issues of community growth as well as decline, and legitimize their status among other communities---in addition to providing a basis for enforcement decisions or conveying community goals and norms. Our findings resonate with work from organizational sociology---including new institutionalism \citep[e.g.,][]{meyer_institutionalizedorganizations_1977, dimaggio_ironcage_1983, tolbert_institutionalsources_1983} and organizational ecology \cite[][]{hannan_structuralinertia_1984, carroll_concentrationspecialization_1985}---which has argued that similarities in organizational structures can be explained by accounting for external social pressures, rather than common internal problems alone. In this work, we extend these insights to the dynamics of rules in online communities, treating them as sensitizing theories to interpret our findings. 

Our paper makes several contributions to social computing research on governance. First, our account of the origins of rules fills a critical qualitative gap in the origins and practices of rule creation and change in online communities.
Using comparative evidence across three different systems, we show how rules mediate online communities' relationships to other communities, platform rules and affordances, and audiences of potential and current members. Second, we argue that a relational account of rulemaking extends prior operational accounts, expanding the design space of digital institutions and community self-governance
by synthesizing concepts from new institutionalism and organizational ecology to explain how rules communicate niche identity and legitimacy. 
Finally, we present a set of design propositions that highlight rule adoption strategies across community lifecycle phases and show how technological design may influence rule adoption and change.

\section{Background}
\label{sec:background}
Governance is the social processes that shape behavior within and between institutions \citep{bednar_governancehuman_2023}. As part of governance, rules are the explicit ``norms, regulations, and expectations that regulate the behavior of individuals and interactions among them'' \citep[p.~5]{march_dynamicsrules_2000}. Although rules are central to governance in online communities, past research takes for granted how rules are initially adopted and how they evolve.
Our paper investigates why and how rules in online communities are created and changed. 

While our theorizing largely happened during and after data collection and is reflected in our Discussion (§\ref{sec:discussion}), we follow influential advice from \citepos{giles_timing_2013} to maintain ``an open mind versus an empty head'' \citep{giles_timing_2013} by conducting a literature review both \textit{ex ante}, to ensure that our work was in conversation with existing theoretical perspectives, and during analysis and writing, to help situate and frame our findings. In particular, two lines of scholarship informed our study. The first includes two paradigms in prior social computing work on online community governance that we call the \emph{social engineering} and \emph{community governance} perspectives. Our research questions and interpretation of findings speak directly to this prior social computing work. We also draw on a second line of work: the new institutionalist and organizational ecology traditions in sociology.
We summarize each in turn.

\subsection{Governance and Rules in Online Communities}
\label{sec:governance}
An extensive body of social computing research has explored rules and governance in online communities---the virtual spaces where individuals engage around shared interests or goals through socially regulated interaction on a social media platform \citep{jones_virtualcommunitiesvirtual_1997, preece_onlinecommunities_2000, leimeister_successfactors_2004}.\footnote{For our purposes, a ``social media platform'' is a website that affords the creation of communities around user-generated content and interactions \citep{carr_socialmedia_2015}.} Two influential perspectives prevail in much of this scholarship.

The first is the top-down \emph{social engineering} approach that centers the role of designers, exemplified by Kraut and Resnick's book \textit{Building Successful Online Communities} \citep{kraut_buildingsuccessful_2012}. This research paradigm seeks to identify generalizable principles by which platform designers can choose rules that shape behavior in ways that correspond to their goals. %
For example, \citet{kraut_buildingsuccessful_2012}'s chapter on norms and rules lists a series of ``design principles'' that offer specific guidance for how a community's designer(s) can use rules to shape participants' behavior.
Many efforts taking this perspective focus on evaluating whether rules successfully mitigate internal problems. For example, research on \textit{content moderation} \citep{grimmelmann_virtuesmoderation_2017} is centrally concerned with the effectiveness \citep{chandrasekharan_crossmodcrosscommunity_2019} and consequences \citep{jhaver_didyou_2019} of rule enforcement practices. 
Other studies test how rule visibility can increase salience and compliance \citep{matias_preventingharassment_2019}, evaluate how post-enforcement rule explanations impact future user behavior \citep{jhaver_doestransparency_2019}, and investigate the effectiveness of different rule enforcement mechanisms \citep{hortaribeiro_platformmigrations_2021,chandrasekharan_youcant_2017}. %
A significant line of research also focuses on describing existing governance approaches and developing new ones that expand how communities can engage in governance processes \citep{geiger_botsbespoke_2014, seering_shapingpro_2017, chandrasekharan_crossmodcrosscommunity_2019, jhaver_didyou_2019, jhaver_humanmachinecollaboration_2019, zhang_policykitbuilding_2020} and content moderation work \cite{grimmelmann_virtuesmoderation_2017}.

The second, \textit{community governance} perspective critiques the first and seeks to understand how end-users create their own systems of governance, even as they may lack full control \cite{kiene_technologicalframes_2019, seering_moderatorengagement_2019, frey_thisplace_2019}. This view highlights a variety of backstage actions and decisions, the complex internal organization of online communities \citep{butler_dontlook_2008,frey_emergenceintegrated_2019}, and the diversity of problems, resources, and goals involved in how communities choose and adopt their rules   \citep{keegan_evolutionconsequences_2017, fiesler_redditrules_2018, nicholson_mastodonrules_2023, reddy_evolutionrules_2023,hwang_rulesrulemaking_2022, frey_effectivevoice_2021, wu_aididnt_2025}.
Scholarship in this vein critiques patterns of ``digital feudalism'' in online spaces, in which online community members have limited opportunities for voice in changing governance of a platform or community \citep{schneider_adminsmods_2022}. Related work has drawn from the Institutional Analysis and Development (IAD) framework created to study common pool resource governance by \citet{ostrom_governingcommons_1990} and colleagues and has drawn attention to higher-order governance layers that include non-operational meta-rules that are focused on rule-making---i.e., a ``collective'' layer for group decision making and the ``constitutional'' layer governing how rules can be changed \citep{ostrom_understandinginstitutional_2005}. For example, \citet{frey_thisplace_2019} argue that improving design at the constitutional layer can make social computing systems more effective and adaptable in changing circumstances.
Overall, the community governance perspective favors localized, community-specific systems and solutions over universal social engineering principles.

Despite their differences, the social engineering and community governance perspectives share a key assumption that most rules %
are primarily operational: rules serve to regulate interactions and contributions \textit{within} a community. Yet, many online communities, particularly those within a platform, have similar rules \citep{reddy_evolutionrules_2023, fiesler_redditrules_2018, nicholson_mastodonrules_2023, hwang_rulesrulemaking_2022}. Both the social engineering and community governance perspectives see rule similarity emerging as communities face similar internal problems and thus select and adapt their rules in similar ways.

Prior empirical accounts of online community rule adoption and change in social computing either adopt the operational assumption or say little about how communities come to have similar rules. Some work explains rule change as moments when rule enforcement creates opportunities to contest the meaning and understanding of rules \citep{delaat_coercionempowerment_2012, crawford_whatflag_2016, matei_wikipediasneutral_2010, kriplean_communityconsensus_2007}, suggesting that communities refine rules to respond to specific internal problems. Observations of rule similarity, for example on Wikipedia, tend to emphasize declining changes and rule persistence, but say little about effectiveness, popularity, or other explanations  \citep{beschastnikh_wikipedianselfgovernance_2008,keegan_evolutionconsequences_2017,hwang_rulesrulemaking_2022,halfaker_risedecline_2013}. %

\subsection{Sociological Perspectives on Organizational Similarity}
In contrast, sociological studies of organizations beyond online communities offer multiple explanations for why organizations adopt similar or identical rules for reasons that are not operational \citep{meyer_institutionalizedorganizations_1977,tolbert_institutionalsources_1983}. 
Although we did not embark on our study with the expectation that these theories would figure heavily in our description of rule adoption and change, we found that they provide an excellent theoretical vocabulary for describing our empirical findings. We introduce key concepts from this literature here that we rely on in our Findings (§\ref{sec:findings}) and Discussion (§\ref{sec:discussion}).

\subsubsection{New institutionalism}
The most important body of scholarship in organizational sociology we rely on is known as \textit{new institutionalism} and was first articulated by \citet{meyer_institutionalizedorganizations_1977} and \citet{dimaggio_ironcage_1983}.\footnote{Confusingly, there are several scholarly traditions in economics, political science, and sociology referred to as ``institutionalism'' and even ``new institutionalism'' with distinct emphases that often define institutions differently \citep{powell_newinstitutionalism_1991}. For example, the IAD perspective described in §\ref{sec:governance} draws from \citet{north_institutionsinstitutional_1990}, \citet{ostrom_governingcommons_1990}, and others associated with what is sometimes called ``new institutional economics.'' This approach defines institutions broadly as ``rules of the game'' and seeks to increase attention to the way written and unwritten rules shape and constrain the behavior of individuals and organizations.
New institutionalism in sociology is a distinct and more conceptually narrow line of work.} New institutionalism argues that organizations' rules and routines do not reflect rational solutions to problems faced by organizations. They are instead based on ``myths'' \citep{meyer_institutionalizedorganizations_1977} and result from a shared environment with isomorphic pressures that cause organizations to adopt similar structures, rules, and routines as a way to earn or signal legitimacy \citep{dimaggio_ironcage_1983}.
\citet{dimaggio_ironcage_1983} argue that organizations will copy routines from other successful groups (mimetic isomorphism), adopt them because they reflect expectations from groups like professions (normative isomorphism), or choose them because outsiders like investors or governments require them (coercive isomorphism).
Although specific sources of isomorphic pressure differ across organizations and environments, this line of scholarship provides an obvious entry point to approach our observation of similarity in online community rules.
After all, online communities on a given platform face similar constraints because they are on the same social computing system \citep{caplan_isomorphismalgorithms_2018}, are subject to similar external policies \citep{tosch_privacypolicies_2024}, and rely on similar technology \citep{hwang_adoptingthirdparty_2024}.

\subsubsection{Organizational ecology}
A distinct tradition within organizational sociology treats organizational behavior as contingent on environmental factors that can promote either variation or conformity \citep{cyert_behavioraltheory_1963, hannan_populationecology_1977}. This \textit{organizational ecology} perspective explains how organizations may differentiate themselves to avoid direct competition and to meet diverse and specialized needs of audiences or markets \citep{carroll_whymicrobrewery_2000}. While the new institutionalists focus on explaining organizational similarity (i.e., why are organizations so similar?), ecologists focus on variation (i.e., why are organizations so different?).
The organizational ecology approach has been echoed by social computing scholars who note that communities can vary immensely in their topics, scales, goals, and norms \citep{chandrasekharan_internetshidden_2018, weld_whatmakes_2021, teblunthuis_identifyingcompetition_2022, zhu_impactmembership_2014, zhu_effectivenessshared_2013,hwang_whypeople_2021}. 
Drawing explicitly from organizational ecology, \citet{teblunthuis_nocommunity_2022} observe how normative differences between communities with similar topics enable communities to be mutualistic rather than competitive with one another. Related work has mapped out ``ecologies'' of the rules and values that would underpin such normative distinctions \citep{chandrasekharan_internetshidden_2018, weld_whatmakes_2021, fiesler_redditrules_2018, nicholson_mastodonrules_2023, hecht_towerbabel_2010}. 
Although researchers have discussed deliberative processes and bureaucratization around rules within specific communities \citep{black_wikipedianot_2008, butler_dontlook_2008}, we know of no work that seeks to explain how ecologies of community rules form in online contexts.

An important distinction between the new institutionalism and ecology traditions is in the way they treat change. 
For new institutionalists, organizational change is a central object of study. For example, a firm may change its rules to comply with a new law or regulation. Similarly, a nonprofit might change its rules to qualify for funding or to conform to published best practices. A key implication from the new institutionalist perspective is that organizations will face similar pressures that will make them structurally similar over time.
Conversely, organizational ecologists rely heavily on the concept of structural inertia, which assumes that organizational change is difficult or even impossible \citep{hannan_populationecology_1977, young_populationecology_1988}. If organizations appear similar, it is because they face similar selection pressures which have caused dissimilar organizations to fail, or because there are positive returns to legitimacy that stem from similar strategies or actions in an environment. That said, some organizational ecologists have argued that certain features of environments may lead to more organizational change \citep{hannan_structuralinertia_1984}.

Although the new institutionalist and organizational ecology perspectives differ in their assumptions and predictions, both perspectives share a key insight absent from most social computing work on governance: organizational forms---including rules---originate to a large extent in the relationships between organizations and their environments. Looking only within communities or at the problems that rules purport to address cannot fully explain how or why rules come about. Instead, both the institutionalist and ecological approaches emphasize the communicative, relational origins of rules. %

\section{Empirical Settings: Subreddits, Fandom wikis, and Fediverse servers}
We study communities drawn from three settings---all sites of prior social computing work on governance and rules. Although most social computing studies focus on a single sociotechnical system at a time, we chose multiple systems to support a comparative account of rulemaking. Looking across settings allows us to consider how the design of systems might impact rulemaking processes. The three systems we selected each include distinct populations of online communities that are created and managed by community members (users): discussion communities (subreddits), peer production communities (Fandom wikis), and federated microblogging communities (Fediverse servers). Below we offer a brief introduction to each system and the ways each is governed.

\subsection{Subreddits}
Subreddits are user-created forum spaces on the networked community platform Reddit.com. A Reddit user creates a subreddit by giving it a name and a topic. Other users may contribute posts (threads to start off a discussion) or comments (to discuss those posts). Subreddits have been created to support a variety of topics, from posting pictures of cute animals to discussing the implications of AI. Reddit users can search for subreddit communities and join by subscribing to a feed that shows the user posts from subreddits they follow. 

Governance of subreddits is primarily enacted by the founder of the subreddit and other volunteer content moderators they appoint \citep{seering_moderatorengagement_2019}. Reddit provides community leaders with a suite of tools, including automated rule enforcement with the AutoModerator bot \citep{jhaver_humanmachinecollaboration_2019}, membership management, and a member communication system called Mod Mail. Subreddit leaders often prominently display their rules on the side of the subreddits' main page and use built-in wiki features to document community guidelines and FAQs. Subreddit rules usually describe topical or behavioral expectations \citep{fiesler_redditrules_2018}. Studies in social computing have mapped out the kinds of rules that are adopted across subreddits \citep{chandrasekharan_internetshidden_2018, fiesler_redditrules_2018} and identified quantitative correlates of rule change \citep{reddy_evolutionrules_2023}. 

\subsection{Fandom wikis}
Wikis are collaboration-based websites where users peer produce public information repositories, like encyclopedias, by directly creating, editing, and organizing content on the website's pages. Fandom is a platform that affords the creation and management of wikis, often made up of members of a fandom who share a strong interest in documenting information about a specific cultural artifact (e.g., TV show, book series, or video game). Interactions primarily occur on the ``Talk Pages'' of articles in a Fandom wiki, where members discuss article content. %
Many Fandom wikis activate a forum feature in the platform.

As the goal of wikis is to produce relevant, high-quality articles documenting their fandoms, leaders frequently develop rules regulating article content. Users who create a Fandom wiki become its \textit{de facto} leaders but frequently appoint other community members to leadership positions.
Social computing scholars have documented how wikis engage in ongoing deliberation about how to change their communities' rules \citep{keegan_evolutionconsequences_2017}, and shown that as wikis grow and mature they become more bureaucratic (by adding more rules) \citep{butler_dontlook_2008} and oligarchic and closed to newcomers \citep{shaw_laboratoriesoligarchy_2014, halfaker_risedecline_2013, teblunthuis_revisitingrise_2018}.

\subsection{Fediverse servers}
The Fediverse is an open, decentralized network of servers (also called ``instances'') that run software compliant with the  ActivityPub protocol. Users join the Fediverse by registering an account on a server of their choice. The server enables its members to share and receive content from other servers, allowing users to interact with fellow server members and the broader network through comments, reposts (``boosts''), likes, and more. We focused on platforms using the Mastodon and Pleroma software, which both offer a microblogging experience similar to systems like Twitter. Servers we observed were generally organized around a shared interest or identity, but members were often not required to post content related to that theme. 

Envisioning decentralized social media ecosystems without centralized owners, Fediverse servers using Mastodon and Pleroma are governed exclusively by the servers' founders, who are referred to as ``admins.'' Admins control membership through open, closed, or application-based user registration; moderate content through admin panels and back-end access to data; and can block other servers to prevent interactions between their server and another (known as ``defederation''). 
Rules on Fediverse servers are usually about speech, especially civility and legality. %
Although scholarly attention to the Fediverse has been limited, researchers have examined migrations to decentralized platforms \citep{dunbar-hester_showingyour_2024, cava_driverssocial_2023, he_flockingmastodon_2023, andradenunes_usersentiments_2023, jeong_exploringplatform_2024}, how platforms conceptualize decentralization \citep{zulli_rethinkingsocial_2020,gehl_digitalcovenant_2023,raman_challengesdecentralised_2019}, community governance challenges \citep{anaobi_willadmins_2023, lacava_polarizationdecentralized_2024, bono_explorationdecentralized_2024, hwang_trustfriction_2025,zhang_troubleparadise_2024}, coordination against hate speech \citep{caelin_decentralizednetworks_2022}, rules on Mastodon platforms \citep{nicholson_mastodonrules_2023}, and the effects of defederation on activity \citep{colglazier_effectsgroup_2024}.

\section{Methods}
To investigate the origins of rules, we treat online communities as the units of analysis in our study. In constructing a sample of communities, we sought variation in terms of goals and platform, but similarities in terms of the contents of rules themselves. We hope that by considering communities with diverse goals and platform contexts, our findings are not entirely driven by either. In contrast, comparing communities with similar rules allows us to assess whether such rule similarity emerges due to shared operational problems or other reasons. 
In order to gather data on the origins of communities' rules, we pursued interviews with community leaders who had participated in the processes of rule selection, adoption, and enforcement. Because we could not identify alternative sources of consistent, archival data on the origins of rules in most communities, interviews offer the sole means by which we can access this sort of information. Reliance on leaders introduces important threats to validity that we discuss in our Limitations (§\ref{sec:limitations}).

We analyze our interview data using an interpretive, comparative approach closely modeled on \citepos{charmaz_constructinggrounded_2015} approach to constructivist grounded theory. Like all grounded theory, the goal of our project is to derive insights drawn inductively from themes in our data. Key details of our methods are described below and are elaborated further in Appendices \ref{appendix:sampling} and \ref{appendix:protocol}.

\subsection{Community Sampling and Participant Recruitment}

Between 2020 and 2021, we built datasets of community rules from subreddits, Fandom wikis, and Fediverse servers. Examples of rules from each of the settings are shown in Appendix \ref{appendix:communities}. Because we were initially motivated by observations of rule similarity across communities,
we computationally clustered communities within each system based on textual similarities between their rules. We then screened clusters to identify a diverse sample of active communities to target for interviews. A more detailed description of this process can be found in Appendix \ref{appendix:sampling}. %

After receiving IRB approval for human subjects research from both institutions involved in data collection, we contacted moderation teams and admins from the identified communities using direct messages on their respective platforms. In total, we recruited 40 community leaders across 34 communities: 9 from subreddits, 14 from Fandom wikis, and 17 from Fediverse servers. Tables describing the communities from which we recruited are provided in Appendix \ref{appendix:community_tabs}.

\subsection{Interviews}
We conducted semi-structured interviews via video calls between June and September 2022. All participants were either volunteer moderators or administrators familiar with their community's rule history and above the age of 18. Interviews ranged from 30 to 120 minutes in length, averaging 60 minutes. A combination of trained undergraduate research assistants and the paper authors conducted the interviews following a protocol that asked participants to describe or tell us stories about their interactions with rules in their communities (see Appendix \ref{appendix:protocol}). We asked questions about the origins of their rules, how they were enforced, when they changed and why, and about related challenges in governing online communities. %

\subsection{Analytical Process}
Our analytical process was based on Charmaz's approach to grounded theory~\citep{charmaz_constructinggrounded_2015}, thematic analysis~\citep{braun_usingthematic_2006}, and collaborative affinity mapping. Small teams of trained research assistants led by four authors conducted line-by-line coding of interview transcriptions using the Taguette software, inductively coding text to break down participants' experiences into actions and feelings through gerunds (e.g., ``relaxing the rules'' or ``feeling a sense of belonging''). Together, we developed over 4,500 open initial codes. Although the vast majority of these were inductive codes that emerged from data, we followed advice from \citet{charmaz_constructinggrounded_2015} and \citet{bowen_groundedtheory_2006} and concepts from the literature discussed in §\ref{sec:background} as sensitizing concepts, resulting in a small number of deductive or semi-deductive codes. 
 
Discussions of the full set of codes across the authorship team led to a rough list of 20 themes. Guided by this list, three authors then conducted a round of focused coding, a step that merges similar initial codes together and elevates them with broader conceptual labels. This led to 337 unique focused codes. 

Three authors then collaboratively engaged in an in-depth affinity mapping exercise with the 337 focused codes to visualize patterns, themes, and processes via a digital whiteboard. Rounds of memo-writing and weekly meetings for discussion and iteration complemented this process. The three authors reached agreement on the most significant contributions to bring forward in this paper: rule adoption and rule change. Finally, the first author engaged in a round of theoretical sampling by using the list of 337 focused codes as a codebook to apply to all 40 interview transcripts. This was to identify supporting evidence across all three empirical settings and to analyze differences when patterns varied between them. The first author distilled a core set of processes and beliefs around rule adoption and change, which was iterated upon with feedback and input from the other authors.

We follow guidelines for qualitative research in CSCW from \citet{mcdonald_reliabilityinterrater_2019} that suggest reporting inter-coder reliability is inappropriate in projects where ``developing codes is part of the process'' or ``where analysis is driven by participants’ own interpretations of their data.'' Similarly, we do not report specific counts of codes. Like all attempts at grounded theory, our goal is the generation of new insight from data, and we leave questions about the prevalence or associations we propose to future work using better-suited data and methods.

\section{Findings}
\label{sec:findings}
From our interviews, we identified seven distinct processes
involved in the production of rules in online communities. The first three processes primarily relate to the adoption of an \textit{initial rule set}. The latter four processes focus on \textit{changing a community's rule set}. %

\subsection{Searching for Rule Templates}
Few leaders we spoke to had imagined a rule set during community creation. Instead, most leaders described the action of \textit{searching for rule templates} in other online communities. For example, FA14, the leader of a toy franchise's Fandom wiki, told us that they ``are always inspired by other communities'' and ``always take a template of rules'' from other communities. Some leaders described feelings of not being ``actually experienced enough and knowledgeable enough'' (FV12) to write rules themselves. FV10, a Fediverse leader for a regional community in the U.S., told us how they ``looked to other examples of rules'' across Fediverse communities during their community's creation to use those rules as templates. FV10 explained:

\begin{quote}
    ``As a lawyer, it's pretty typical to look at other example documents if you're writing a contract or something. You look at other examples of how people have done it...and that's kind of how our rules developed. You see what challenges other people have encountered, and then you can assess whether or not that's a challenge that you think that you need to be prepared for.''
\end{quote}
FV10 rationalized the practice of \textit{searching for rule templates} by comparing rule adoption in online communities to policy writing in modern day firms. FV10 showed how \textit{searching for rule templates} can expose leaders to a range of possible challenges they may anticipate and also provides them with examples of rules to overcome those challenges that they could then choose to adopt.

\subsubsection{Regulating Content and Behavior} 
Leaders often cited two operational motivations for adopting rules: \textit{regulating content} and \textit{regulating behavior}. While these insights align with a large body of work in past social computing scholarship about the importance of regulating content and behavior in online communities \citep{kraut_buildingsuccessful_2012, grimmelmann_virtuesmoderation_2017, gillespie_custodiansinternet_2018}, we find our participants' concerns worth reporting given the different contexts for their needs. %
Subreddit leaders discussed a variety of rules regulating content, such as rules for meaningful discussions (R1), rules restricting advertising (R4), or rules around sharing ``misinformation'' (R5).   
Fediverse leaders mostly discussed a need for rules that restrict content they believed was inappropriate for professional work settings (NSFW content), ``hateful'' content, advertising, and spam. Regulating content for Fandom leaders was important for setting standards and quality control for their wikis, like how information is presented as well as accuracy (FA6). 

\textit{Regulating behavior} was an important justification for rule adoption across all three settings. For subreddit leaders, this was expressed by a common sentiment that the purpose of rules ``is to keep everybody safe'' and ``make it so the community doesn't become a toxic place'' (R9). Although Fandom leaders placed less emphasis on facilitating discussion, leaders still adopted rules to promote civility in Talk Pages, discussion forums, and in affiliated Discord servers that many Fandom communities used for synchronous chat. For example, FA11 explained how their rules are ``there for safety'' and ``to make people feel like they're included and they belong.'' Behavioral regulations were especially important for Fediverse leaders, many of whom had migrated from Twitter to the Fediverse due to concerns about ``a lot of abuse and harassment'' going unpunished (FV8) and the sense that Twitter had become ``too toxic'' (FV7) to use.
Fediverse leaders reported adopting rules to protect their members from specific kinds of harassment and hate speech, such as racism or transphobia (FV16). %
In sum, leaders chose initial rules for operational reasons:  to establish expectations of acceptable behavior and content and in turn to structure participation to achieve their communities' goals.

\subsection{Copying Other Communities' Rules} \label{copying}
One of the original questions motivating our study was ``why do so many online communities have similar rules?'' In the previous section, we showed how leaders would seek out other communities' rule sets to use as starting templates. For many, the next step was to copy. Copying was most evident in our interviews with members of the Fediverse. FV10 explained this step as ``copying the good stuff'' after they ``look[ed] at other examples of how people have done it.'' Leaders would copy and paste rules they felt were a good match for their community and then later ``sort of build on it'' (FV10) with further customization to their communities' unique needs. That said, some leaders from the Fediverse said they copied the \textit{entire} rule set of another community and never customized or changed it. In explaining these actions, these leaders emphasized a relational function of rules with the goal of \textit{establishing external legitimacy} among other communities in their decentralized networks.

\subsubsection{Establishing External Legitimacy}
A key motivation behind the process of \textit{copying other communities' rules} was the desire to project legitimate governance (i.e., the presence of responsible content moderation) and communicate alignment with established norms across an ecosystem of online communities. Fediverse leaders have the power to block (``defederate'') communities that threaten their well-being and goals. These same leaders felt their choices in rules were being judged by admins of other Fediverse servers who they imagined  could defederate at some point in time. FV10 said this adds ``an important function of rules'' because it ``sends signals to other admins that they know what we are moderating.'' This external scrutiny from other admins motivated Fediverse leaders to adopt rules because these rules would signal accountable governance and content moderation. In this way, rules help \textit{establish external legitimacy} for communities in decentralized networks.

Many Fediverse leaders pointed to \texttt{mastodon.social}'s rule set as a benchmark of legitimate community governance. \texttt{mastodon.social} was one of the first communities on the Fediverse and was created by Eugen Rochko, the developer of the most widely used platform for hosting Fediverse communities (Mastodon). Fediverse leaders often looked to this community because it is ``the largest instance'' and it is ``hosted by Eugen'' (FV5). FV5 explained: ``You know, if you're gonna copy rules about a new community, you might as well go right from the source. So I copied most or all of the rules [from \texttt{mastodon.social}].'' FV2 came to a similar conclusion:
\begin{quote}
 ``I went to other instances. I looked at what they had in terms of rules, and the rules from the \texttt{mastodon.social} instance made sense to me. And I also saw a lot of other instances referring to those rules, which kind of created this network effect.''
\end{quote}
FV2 observed many other communities on the Fediverse drawing on \texttt{mastodon.social}'s rule set, which helped rationalize its adoption as well. FV2 described a self-reinforcing pattern of rule adoption where communities added rules to signal their alignment with the standards and norms of a reputable, established community. 

In addition to borrowing from \texttt{mastodon.social}, several Fediverse leaders integrated the Fediverse-Friendly Moderation Covenant (FFMC) into their rule sets as a pledge to adhere to shared, cooperative standards of governance with other Fediverse admins. Adopting the FFMC not only signals a commitment to cooperative governance between servers but also aligns a server with a ``bubble of known instances,'' establishing the admin as ``a stand up citizen, a stand up member of the Mastodon community.'' (FV1) These alignments, whether they are achieved by adopting \texttt{mastodon.social}'s rule set or pledging to the FFMC, could help build trust among peer admins and could lend emerging communities on the Fediverse the external legitimacy they believe is needed to stay in the network.

Subreddit and Fandom wiki leaders did not mention copying other communities' rules purely to signal legitimacy in the way that Fediverse leaders did. That said, some Fandom wiki leaders like FA3 sought out rules from ``other big wikis'' to see how others addressed similar issues and used their rules as starting points. Subreddit leaders like R4 and R5 mentioned copying guidelines from the ``Reddiquette,'' which, like the FFMC, is an informal set of behavioral guidelines written by other Reddit leaders. 

\subsection{Customizing the Adopted Rule Set} \label{customizing}
Customizing the adopted rule set was a natural next step for many leaders. After all, the communities from which they drew their initial rule sets often differ in their specific goals. Interviewees from Fandom and Reddit offered insight into this process. FA1 %
explained that ``it would be nice'' if every other Fandom wiki had the same rules, but that rules have to be modified to each community's unique needs: ``every community is different.'' Leaders often searched for specific kinds of rules from different communities to solve their problems. FA4 %
described this process in the context of creating a rule on image usage by customizing the image policy from another wiki: 
\begin{quote}
``We didn't just copy the rules. We saw exactly what those other communities were asking of their users---and we saw what worked for us, what didn't, and what the other sites were not asking. And then we adjusted to our specific community.''
\end{quote}
We note that changes were not driven purely by differences in community topic. Leaders also described trying to improve the language or effect of a rule. For example, R4 based their rules on other communities they had led and made them ``better to a certain degree.'' Sometimes leaders described these edits to other rules as minimal and changing them ``a little bit.'' (FV12) Other times, the process of rewriting the rules could be ``agonizing'' (FV11) as leaders struggled to make their rules ``well-written.'' %

\subsubsection{Reinforcing Community Identity} 
The process of \emph{customizing the adopted rule set} revealed another relational purpose of rules: \emph{reinforcing community identity}. While rules that structure community identity fulfill an operational purpose by restricting contributions that don't achieve community goals, they also serve a relational purpose by setting a community apart from others that overlap in membership or topic. This was especially important for subreddits and Fandom wikis, where overlapping communities attract audiences who may arrive with misaligned expectations from their experience in similar groups. For example, R4 adopted a rule requiring that contributions ``must be relevant'' to IT professionals because newcomers would arrive asking for IT troubleshooting. %
Rules like these helped to make explicit the community's intended audience (for example, ``technical professionals'' for R4's community), which helps reinforce community identity \citep{kraut_buildingsuccessful_2012}. We also found that leaders may try to reinforce their community's identity by explicitly directing newcomers to other communities that have been partitioned to fulfill a need, such as R4 creating and linking to an offshoot subreddit specifically designed for IT troubleshooting help.

In another example, R2 manages two subreddits: both involved in Q\&A style discussions, but they differ in the level of formality and seriousness. R2 explained that the rules in the first community have to be strict and structure a ``rigorous environment'' while the rules in the second community allow for more freedom in what can be asked and how it can be answered, noting that the rules help keep that community's discussions ``as open as humanly possible.'' R2 said ``We [at second community] want a place that's not [first community].'' By establishing these distinct expectations with their rules, both communities thrived despite overlapping in a similar niche space. The rules helped distinguish the communities by the style of Q\&A format: one for formal and serious discussions and one that's completely informal and casual. The tailored rules communicate to both members and newcomers not only what the community is, but also what it is not.

Rules were reported to help reinforce identity in Fandom communities as well. For example, FA9, a leader for a comic book franchise, compared their rules to the rules of the wiki for a different comic book franchise to explain how their community was unique. While both comic book wikis utilized a forum feature, FA9 noted that their wiki's forum rules promote ``meaningful discussions'' by banning conversations about fan fiction, while the wiki they were contrasting theirs to has more ``permissive'' rules. FA9 lamented the disruption from newcomers expecting rules like those in the different wiki community, telling us that their wiki is ``not the place'' for that kind of content. Like R2, FA9 used rules to explain what their community is and isn't about.

While some Fediverse communities portrayed an identity (e.g., a Mastodon server for people in a specific city in the US or a community for amateurs of a specific hobby), their rules usually did not limit discussions based on their community's identity. %
However, some Fediverse leaders saw the absence of rules restricting discrimination in other communities as an indicator that they allow hate speech to flourish. These were called ``free speech'' servers, and it was inferred that content moderation was virtually absent. FV13 said leaders of free speech communities ``don't care about moderation---[they let you] say what you like.'' FV8 told us that they adopted their strict rules on discrimination and hate speech to ``signpost'' to outsiders that their community is not a place for ``free speech warriors.'' In one instance, FV14 said their rules essentially help people self-select out because registration requires them to read and agree to their anti-discrimination rules.

\subsection{Encountering Problems that Challenge the Rules}
Not all communities felt the need to change rules---particularly those closed to new members or low in activity.  For FV17, an absence of problems in their community resulted from the tight-knit and homophilous nature of the group who ``kind of know each other already.''  Notably, on the Fediverse, leaders can close account registrations on their server and thus control who may become a member. For example, FV10's server ``pretty much never had open registration'' and thus, the admin explained, never had to change rules. Similarly, relatively small or new communities, like FA6's wiki community, rarely saw problems with rule breaking because ``the number of edits was in the very low volume.'' When membership and activity are low, there may be fewer opportunities for problems to emerge and threaten community goals. That said, many leaders reported changing their rules in response to problems in their communities. %
Our interviewees discussed how a community might encounter problems and then change its rules in response---particularly as communities grew or declined. 

\subsubsection{Problems from Growth} \label{growth}
Leaders often considered whether the community was able to achieve its goals under its old rule set when their communities grew. In particular, %
content-related issues tended to emerge alongside growth and occasioned leaders to reevaluate and reinforce their communities' goals by making rules \textit{controlling quality and identity}. For example, FA1 told us the main goal of their community was ``to document and be a reliable resource for people learning about different trends and communities in real life and online.''
When an influx of new people created articles not about \emph{trends} but rather about their own unique personal style, FA1 said they had to ``crack down'' on %
this problem by
implementing a rule that restricted the creation of article pages based on one's own personal style because ``multiple pages for things that don't exist chips at the reliability.'' %
Reliability is an important feature of many wikis, but allowing the creation of articles that aren't grounded in fact might jeopardize the goal of being a reliable source of information.

Community growth also tended to coincide with the addition of rules \textit{restricting spam}. FV5 explained how in their Fediverse server for musicians in a large city, they added a rule to control spam by self-promoting local artists explaining: ``I didn’t want the timeline to simply be a sort of clogged artery of promoters and no people actually talking to each other.''
Even though the problem was caused by members acting in good-faith, FV5 felt that too many such posts threatened the community's goal ``to be a space for people to talk to each other.'' Similarly, R5's community %
created a new rule to limit posts about topics that gain sudden virality and importance because they can set off a cascade of members posting threads with identical topics. %
R5 explained their rationale for rulemaking to prevent such spam during the COVID-19 pandemic:
\begin{quote}
    ``Everyone was just losing their shit posting stuff every, I don’t know, 15-20 seconds at some point. I mean, it got ridiculous. So we set a rule in place that, okay, ``Mega Thread Mode'' is active. So what that essentially means is that there’s one thread at the top [of the subreddit’s webpage]---it’s stickied. Anything related to that subject gets pulled into that thread.''
\end{quote}
Implementing a rule restricting posts about suddenly important topics to a single ``Mega Thread'' helped to reduce the noise that could result from a sudden influx of people posting the same news articles and allow space for other discussions to emerge.%

\subsubsection{Problems from Decline} \label{decline}
Leaders also attributed rule change to community decline, which might manifest in \textit{diminishing member activity} or \textit{neglecting members' problems}.  Some leaders responded to periods of diminishing member activity by changing the rules to \textit{reduce barriers to participation}. FA5 reported that members had ``complained'' on Talk Pages about ``feeling kind of suffocated'' by some rules. FA5 said that leaders responded by making  requirements less restrictive, believing that they ``need to not be so hardline on handling these things.'' FV5 told us how they hadn't yet changed rules but ``might have to take rules away if they seem overly restrictive.'' 

A more extreme example of changing the rules due to diminishing member activity comes from R3's community, a place to discuss a television series. Following the show's conclusion, activity in the subreddit began to plummet. According to R3, their community's moderators responded by evaluating not just the rules, but the community's identity:
\begin{quote}
    ``We had a discussion about---okay, are we going to relax the rules? How far are we going to relax them? Are we going to allow people to post fan-fiction? Or even allow people to post memes?'' 
\end{quote}
``Relaxing the rules'' to allow fan-fiction and memes would open up new opportunities for members to make novel contributions to the community and sustain activity that isn't solely rehashing discussions about the show's content. But the result would also mean the community would become a more casual space from the proliferation of fan stories and low-effort memes. 

\paragraph{Changing Leadership to Manage Decline}
In rarer cases, communities were reported to recover from \textit{neglecting members' problems} in the rules by radically \textit{changing leadership}. For example, FA12 told us that the previous team of admins for their Fandom wiki was changed for being ``lax with users who were being discriminatory.'' %
Harassment on Talk Pages went unaddressed by moderators, leaving members feeling unsafe. The problem was eventually addressed through a ``big overhaul of the moderator team'' who also updated the rules to crack down on discriminatory language.  In R9's community, where members exchange goods through the postal mail, it was claimed that the previous leadership was ``pushed out because they weren't putting in any rules to protect people from scams.'' Members did not feel safe and participation declined as the leaders failed to protect them. R9, along with some other core members, became their community's new leaders through a community-wide vote and immediately added rules requiring verification photos, commenting on posts before privately messaging, and using PayPal Goods and Services for trading physical goods. 

\subsection{Becoming Aware of the Need to Change the Rules} \label{becomingaware}
While rules addressing internal problems were primarily operational, we found that each stage in the processes of changing rules has significant relational aspects. The first stage of a rule change is always becoming aware of (new) problems so as to recognize the need to change, add, or delete a rule. Here we elaborate two ways community leaders became aware of problems.

\subsubsection{Reporting to Leadership}
Community leaders often learned about problems not through direct experience but from fellow leaders or community members. Leaders in FA1's Fandom wiki, having recognized a problem, would ``always discuss it with the [other] moderators,'' to decide on whether they should implement a change. Leaders sometimes relied on members for awareness of problems.  For some Fandom leaders, rule change processes would happen ``whenever the user base complains'' (FA11), in which ``someone has an issue with a rule, they bring it up---on the forum---to an admin'' (FA3). Subreddit leaders learned about problems from members not only through direct messages or the ModMail system, but also in ``meta'' threads for conversations about the community.  Some subreddit leaders said they openly encouraged such feedback to learn how to improve governance. Still, some problems were acknowledged by leaders but left unaddressed.  

\subsubsection{Reflecting After Enforcement}
Interviewees discussed how everyday practices of community governance could prompt reflection about the efficacy of having rules. For example, FA3 said that their community lacked rules in the early days, calling this time ``the wild west.'' Only after noticing an increase in problems related to members' behaviors and quality issues in content did they feel it necessary to adopt a rule set.
While FA3 highlights an example where new rules were motivated by the absence of rules, R8 noted how the enforcement of existing rules could reveal ways the current rules were insufficient and lead to discussions concerning community management:%
\begin{quote}
    ``At some point, we sit down and decide that the rules we have aren't clear enough or that they don't probably reflect the way that they have been enforced. And so we try to rewrite them all just based off of how we actually have been applying them.'' (R8)
\end{quote}

These moments of reflection apparently involved formalizing and streamlining norms. FA4 explained that a reason for one of their biggest rule changes was ``to overhaul'' their communities' rules to ``properly consolidate all the unwritten rules.'' Leaders also said they would overhaul rule sets when they became bloated by rules that were redundant, unnecessary, or hard to enforce. FA12 explained they overhauled rules because their existing rule set was ``really complicated and kind of difficult to keep track of.'' In sum, sweeping rule changes seemed to occur in mature communities where inaction would threaten community goals.

\subsection{Deliberating on How to Solve Rule-Related Problems} \label{deliberating}
Leaders often discussed deliberation processes with other moderators and admins to better make sense of new problems, what the current rules do to combat those problems, and how possible rule changes might ultimately resolve them. %
Leaders on Fandom and Reddit often convened in group chats using software such as Discord where ``everybody [is] adding their own [two] cents to it and then changing [rules] as a group'' (R7). FA1 described a situation where their team of moderators had a three-hour voice call reviewing and ``cleaning'' up their rule pages. FA4 similarly recalled a ``lengthy process with the rest of the administrative team'' when they drafted new rules for their community. Leaders described various procedures for these discussions. For example, R9's group established discussion norms to make sure the other leaders ``all get along well'' so that they can ``communicate disagreements and not have it be a big thing.'' When disagreements did occur, the leaders were ``respectful enough that they'll get to a point where one of them is swayed one way or the other,'' and if they couldn't, then the issue would be ``tabled for another month or two.'' %
FA6 described a procedure for teasing out potential issues with a new rule:
\begin{quote}
    ``When we propose a rule or we're thinking through it, I tried to test it for edge cases---like the gray areas---to see how we [handle] that or whether we have to change the rules or whether it fundamentally makes it not very enforceable.''
\end{quote}
In this case, leaders considered the design of proposed rules and how feasibly they could be enforced by testing them against imagined situations where breaking the rule would not be totally clear.%

Some leaders valued \textit{including community members} to seek information and feedback on community problems and solutions.  Members' experiences mattered to leaders who didn't want to implement rules that might disrupt their activities. FA6 called this ``doing outreach'' to understand what members wanted and how they might react to a rule change. Discussions happened in spaces such as ``Meta'' threads on subreddits and discussion forums or Talk Pages on Fandom. FA3 described this back-and-forth deliberation with members and leaders:
\begin{quote}
    ``We try to get input from other users, and if we come to a decision and everyone else is like 'No, that decision sucks', then we're like `Well we think it works better because this, this, and this, but we see your point and we can compromise here, here, and here on that change.''' 
\end{quote}
In this example, FA3 would work collaboratively on changing rules by finding compromises between the goals of the leaders and the experiences of community members. R4 described their approach to this procedure of rule change as  ``consultation with the community.'' Community members would make problems visible to leaders by ``flagging posts'' and ``complaining'' to moderators about inappropriate behavior they witnessed. R4's leaders would communicate with members about their intended rule changes to ensure the changes wouldn't disrupt members' experiences. Leaders on Reddit said they often used the ``upvote'' feature, which allows community members to vote on particular comments within meta discussion threads, to help understand ``what people were wanting'' (R5).

\subsection{Making the Decision to Change the Rules} \label{makingdecision}
After encountering and observing problems and deliberating over potential solutions, leaders would decide whether or not to change the rules. Decision-making processes varied within each setting, but we observed some patterns: 1) autocratic decision-making for leaders of Fediverse communities, 2) oligarchic decision-making for leaders on Fandom, and 3) a (limited) implementation of democratic participation for rule change decision-making in some Reddit communities. 

Fediverse leaders we spoke to \textit{rarely} changed their rules. When rules were changed at all, they were mostly revised ``to make them more clear'' (FV11). Most problems they mentioned originated in external servers and were resolvable through the Fediverse's ``defederation'' system of blocking other servers from interacting with their communities. Internal problems were reportedly rare because new members were often vetted in a registration process to join. Additionally, Fediverse leaders often felt like they deserved to govern in a seemingly autocratic fashion, sanctioning rule breakers without worrying about backlash. FV4 explained their reasoning:%
\begin{quote}
    ``Ultimately... I'm paying to run the server. I don't really wish to pay to give someone who's been flagrantly awful a platform. They can go build their own soapbox. I don't wish to use my time and money and resources to give them a broader audience.''
\end{quote} 
FV4 highlights an arrangement that's common for Fediverse admins but absent for leaders of subreddits and Fandom wikis: they must shoulder the responsibility and operating costs of hosting the community's web server. Consequently, they felt this granted them more control over their communities' governance. Rather than join a community to challenge their rules, Fediverse leaders felt like members are expected to know and respect ``the rules and the norms of operation...for a community %
or treating joining a Fediverse community like you're ``engaging with a contract with a company'' (FV13). As a result, the rules often reflected what leaders ``want personally in a community experience or social media experience.'' (FV1)

Deciding to change the rules was often complex for communities on Fandom and Reddit. In both platforms, communities were typically governed by more than one person. In Fandom communities, decisions would be made within ``a small collective of moderators and admins'' (FA3). Leaders said they sought consensus on how to handle problems in discussions on changing the rules: ``You always try to compromise or try to make your interests---make your points---to have these rules [or] not have these rules'' (FA14). Hierarchical decision-making structures were reported even within these small groups of leaders. FA1 pointed out that the original founder of their wiki community changed their rules with no deliberation from other admins when they came back from a six-month hiatus because they ``decided that the rules were too lax.'' Fandom leaders had formalized roles, including bureaucrats (who held the most decision-making power), admins (who helped manage community-wide problems), and moderators (who did much of the rule enforcement). In FA11's wiki, the ``main bureaucrat'' is ``the final decision maker'' who has ``the central role of helping make the rules for the wiki and making the final decisions if there was a tie'' when voting on changes.

While subreddit communities also made governance decisions within their small team of volunteer moderators, we heard of cases in which leadership felt compelled to involve community members in the process. In the previous subsection, we described the ways in which R4 believed ``the good way to handle rule development is based off the community's needs and [to] consult them the whole way.'' R4 added rules ``through a process of doing surveys every quarter'' for their community in which they identified problems and asked members, ``what do you think we can do to fix this?'' Answers to the survey would be taken as suggestions to ``build those rules'' that would improve the community. %
In one example, R4's team sent out a questionnaire giving members several options for how leadership should handle advertisements from corporate blogs in their community:
\begin{quote}
    ``We suggested a couple options that we're like `Just remove them all' or `Keep them but flair\footnote{Flairing a post means that the title of the discussion post is labeled with a visible tag.} them and mark them as advertising', that kind of stuff. And that ended up being the result that we decided was probably best for the community. That's forever another rule.''
\end{quote}
R4 claimed this was the best approach for their community because they felt a need to appease the high-quality contributing members on whom they relied.

When R9 and others replaced a previous generation of leadership after a period of neglect, they adopted an approach that incorporated community members into the decision-making process for changes. R9 explained that they do ``community posts'' once a month to allow members to give feedback about their experiences. When reviewing the feedback, if R9 sees ``people a little bit cranky about some rules,'' they will ``put out a poll...to see if they [members] want things to change or not.'' Members vote on whether a change should be implemented, although sometimes, R9 told us, ``many polls...end up basically being 50/50.'' They adjourn and do another poll a month later. In this way, R9's leaders ensure that they ``change things as the community wants to be changed,'' in order to make sure the community is going in the right direction, supporting community members' goals, and safeguarding their experiences.

\section{Discussion}
\label{sec:discussion}

Although leaders often adopt rules to regulate behavior and content, we find that this is only part of the story. Rules also reinforce a community's distinctive identity within an online platform and establish a community's legitimacy in broader networks. This insight complements and challenges prior operational accounts of the sources and purposes of rules in social computing work, from both social engineering and community governance perspectives, and reveals underappreciated relational aspects of rulemaking. Some of these relational aspects are external, such as when rules draw distinctions or signal alignments between communities. Others are internal, as when rules reinforce identity or consensus among a community's leadership team or between leaders and regular members. Both internal and external relationships are important in the processes of rulemaking; they draw attention to problems, help form solutions, and shape decisions about changing the rules. Our work also extends prior quantitative research in CSCW on changes in community rules \citep{frey_emergenceintegrated_2019, reddy_evolutionrules_2023} by showing how leaders think about rule creation and change, as well as how different community platforms may shape rulemaking processes by %
constraining relations in different ways.

In the rest of our Discussion, we consider the implications of relational dynamics of rules in governance systems for online communities. We then reflect on how our findings inform theories of institutional design and change in online communities %
as well as directions for future research on community governance. %
Along the way, we highlight a set of design propositions that suggest future work---both system building as well as empirical evaluation---derived from a more relational understanding of online community rules. We present these as tentative propositions for future design and evaluation work to illustrate the applicability of our theoretical framework, rather than as definitively proven guidelines.%
    
\subsection{Relational Dynamics of Online Community Rules}
A prevailing assumption in prior scholarship is that rules formalize social norms that regulate interactions \textit{within} communities \citep{butler_dontlook_2008, kraut_buildingsuccessful_2012,grimmelmann_virtuesmoderation_2017, chandrasekharan_crossmodcrosscommunity_2019, seering_buildingmore_2019}.
Rules are treated as instruments for directing human interactions in computer-mediated environments. We find that, in addition to advancing this operational purpose, rules also serve relational purposes as leaders design rules with respect to other communities, leaders, and users. %

First, rules set and sustain community boundaries, particularly to protect community identity amid evolving membership dynamics. The challenge of protecting identity is important---particularly for communities prone to influxes of new members who may alter the community's composition or undermine its focus \citep{lin_betterwhen_2017}. In response, communities discoverable on large platforms and open to broad participation often implement rules that restrict off-topic content. While prior work has suggested that rules contribute to identity formation \citep{butler_dontlook_2008, fiesler_redditrules_2018, seering_moderatorengagement_2019}, our findings show that communities may defend their identity over time by prescribing strict guidelines for participation, often through reference to related communities. We extend prior work by \citet{hwang_whypeople_2021} and \citet{teblunthuis_nocommunity_2022} who emphasize the importance of membership size and niche audiences within ecosystems of interdependent communities. This relational use of rules helps explain the concept of ``nested niches,'' in which communities thrive through their embeddedness in a broader socio-informational ecosystem \citep{hwang_whypeople_2021}. Rules may provide tools for drawing and maintaining boundaries between specialist and generalist spaces---especially on platforms, such as Reddit, where communities often emerge from one another and share overlapping members and topics \citep{tan_tracingcommunity_2018, teblunthuis_identifyingcompetition_2022, teblunthuis_nichedynamics_2025}. %

\vspace{10px}
\begin{mdframed}
\textbf{Design Proposition:} Community leaders may improve rule understanding by conveying how the rules define their community identity and what makes them distinct from others \citep{matias_preventingharassment_2019}. 
Designers of social computing systems can support this by creating ``introductory spaces''---e.g., side panels, about pages, pinned messages---that help newcomers understand what differentiates the community. 
Platforms might also provide %
leaders guidelines for customizing their rules to reflect the community's position within the platform's ecosystem. 
\end{mdframed}
\vspace{10px}

Second, rules may signal legitimacy to potential newcomers and leaders of other communities. Doing so can demonstrate commitment to governance as a means of attracting newcomers and establishing a good community reputation. %
This pattern was most salient on the Fediverse where reputation is vital for maintaining connections to other servers. Many Fediverse leaders adopted the same rules as the most popular servers because these were a widely recognized standard for governance. A community's rules thus signal to important external audiences that the leadership takes governance (i.e., content moderation) seriously and abides by shared social norms. 

This finding relates to recent work on Fediverse rules proposing that the decentralized arrangement of Fediverse platforms leaves no higher authority to hold individual communities accountable, which may enable toxic and abusive online cultures \citep{nicholson_mastodonrules_2023}. While Fediverse servers \textit{could} adopt heterogeneous rules, isomorphic pressures from the surrounding environment \citep{dimaggio_ironcage_1983} push server leaders to adopt identical rules more than the other settings in our study. 

Such institutional isomorphism is dramatic on the Fediverse, but it reflects an observation from organizational sociology. Leaders of new online communities, facing uncertainty in how to succeed in their environment \citep{singh_organizationallegitimacy_1986}, are prone to failure \citep{kraut_buildingsuccessful_2012} and may lack requisite resources or knowledge to write well-crafted rules on their own. They might model their community's institutions (including their rules) on communities that they perceive to be legitimate or successful \citep{dimaggio_ironcage_1983}. In our study, founding leaders believed that a robust rule set would both attract newcomers and strengthen relations with neighboring communities. 

\vspace{10px}
\begin{mdframed}
\textbf{Design Proposition:} Community leaders %
often seek legitimacy by aligning with broader, recognizable norms across the platform.
Platforms can facilitate this %
by %
providing curated rule templates drawn from similar or neighboring communities.
\end{mdframed}
\vspace{10px}

Our findings also point to ways that sociotechnical system designs can condition rulemaking. Rationales around legitimacy were paramount on the Fediverse but largely absent from interviews with communities on Fandom and Reddit. Meanwhile, rationales around identity were central on Reddit and Fandom but less salient in interviews with Fediverse leaders. We hypothesize that this difference is attributable to how communities on Reddit and Fandom are strongly held to platform-wide norms and policies, while Fediverse servers depend on connecting to a decentralized network of autonomous communities. Platform-wide norms reduce the uncertainty that creates isomorphic pressures, making the problem of identity differentiation more central. By contrast, because Fediverse communities can engage in community-level blocking \citep{colglazier_effectsgroup_2024}, the stability of a decentralized network may depend on leaders aligning on some broad set of norms and rules regulating inter-community relations.
No similar affordance for one community to sanction another exists on Fandom and Reddit.

\vspace{10px}
\begin{mdframed}
    \textbf{Design Proposition:} Leaders of communities in decentralized or federated networks may seek normative alignment through cross-community agreements or covenants to stabilize relations with other communities.
    Designers of these systems where community-level sanctions exist should create tools that help communities coordinate normative alignment.
\end{mdframed}

\subsection{Navigating the Challenges of Rule Change}
Recent work in social computing has emphasized that users, despite being members of online communities, frequently lack effective mechanisms to voice concerns and implement changes that would improve their experiences ~\citep{frey_effectivevoice_2021, schneider_adminsmods_2022}. Such deficiencies stem from sociotechnical system design where the single founder of an online community on most platforms is given autocratic control over all aspects of community governance. Under this arrangement, change depends on the founder, who often becomes a ``benevolent dictator for life.'' %
Our findings reveal some ways that community-driven change is possible under this arrangement and how both leaders and members may work together to resolve community problems. 

First, additional problems may trigger a need to change the rules in an online community. Recent quantitative studies of rule changes in subreddits and Minecraft servers have found that increasing membership size and community age are associated with leaders deciding to change the rules \citep{frey_emergenceintegrated_2019,reddy_evolutionrules_2023}. We find that periods of decline, such as decreasing activity in the community, may also provoke rule changes. However, instead of adding rules to restrict content, leaders may remove or relax rules to encourage wider participation. 

\vspace{10px}
\begin{mdframed}
    \textbf{Design Proposition:} Rule changes may stem from periods of participation growth as well as periods of decline. 
    In times of flux, platforms could prompt leaders to re-evaluate their goals as well as their rules \citep{reddy_evolutionrules_2023}. Rapid growth may provoke rule expansion and clarification \citep{kiene_survivingeternal_2016}, while rapid decline may signal a need for broader pathways for participation.
\end{mdframed}
\vspace{10px}

Next, we find that leaders, while knowing problems exist, may lack the capacity to implement effective rule changes to resolve those problems. \citet{march_dynamicsrules_2000}, analyzing the evolution of written rules at a university, argue that organizations are ``thermostatic'' in their attention to problems: as long as performance is satisfactory, problems may go unresolved or unnoticed. Self-governed online communities may be especially prone to neglect problems because their leaders are typically unpaid \citep{dosono_moderationpractices_2019, matias_civiclabor_2019, wohn_volunteermoderators_2019} and frequently quit because of their lack of time, which brings about psychological distress \citep{schopke-gonzalez_whyvolunteer_2024}. Yet, as communities mature, new problems often demand new rules \citep{butler_dontlook_2008, kiene_survivingeternal_2016, lin_betterwhen_2017, reddy_evolutionrules_2023}. Leaders we interviewed cited similar challenges in finding time to change their rules, carefully weighed the demands of potential new rule enforcement, and in one case, %
replaced the leadership team with members better positioned to improve community governance.

\vspace{10px}
\begin{mdframed}
    \textbf{Design Proposition:} Volunteer leaders may face time and attention constraints that impact their capacity for proactive governance. 
    Platforms could provide tools that 1) help recruit and distribute lightweight governance tasks to trusted community members, and 2) surface and act on member feedback via structured feedback mechanisms like surveys or binding community votes \citep{frey_effectivevoice_2021}. 
\end{mdframed}
\vspace{10px}

One way to reduce strain on leaders could be to share governance work with the broader community. Doing so taps into community members' collective knowledge to help in identifying problems and co-creating solutions \citep{turco_conversationalfirm_2016, hwang_adoptingthirdparty_2024}. 
None of the communities we investigated for this study had binding mechanisms through which community members' voice directly influenced decision-making by leaders \citep{frey_effectivevoice_2021, schneider_adminsmods_2022}. Fediverse leaders we interviewed were particularly autocratic: they rarely changed their rules, never held meta discussions or encouraged feedback, %
and made decisions with no input from members. In fact, Fediverse leaders often said they wanted to create their \textit{own} server where they would get to decide how things should be. In contrast, Fandom leaders expressed concerns about managing delicate relationships with and gathering input from valuable contributors. Subreddit leaders seemed even more involved with and accountable to their members. Many reported rulemaking processes that involved much more participation from members.

What makes Reddit so different from the Fediverse? We propose that a combination of sociotechnical affordances and institutionalized norms sustains a stronger culture of participatory governance. On the affordances side, centralized discussion spaces and integrated voting tools help members collectively communicate their concerns and values to leaders, increasing pressure on them to solve problems. On the normative side, Reddit users have developed sitewide values emphasizing individual freedom and skepticism of moderators who act unaccountably. While nothing binds subreddit moderators to accept member demands, soft binding mechanisms, along with low costs of exit, might constrain moderators. For example, several subreddit leaders we interviewed were cautious about making decisions without input from their community members. They feared that if influential members' concerns weren't taken seriously, these members would take their expertise elsewhere. Similar examples with Fediverse leaders were non-existent in our sample.

Taken together, these arrangements in Reddit may result in mutually dependent leader-member relationships: members need leaders to moderate content and enforce rules, while leaders need members to contribute content and sustain community activity. Evoking metaphors in content moderation from \citet{seering_metaphorsmoderation_2020}, the leaders of these subreddits may be more like ``governors'' who satisfy the needs of their constituents and less like the ``custodian'' wiki leaders who curate and tidy wikis, or the Fediverse ``dictators'' who create and run servers however they see fit.

\section{Limitations}
\label{sec:limitations}
As with all attempts to build a grounded theory, our findings are offered as %
a set of untested and unproven claims.
These claims are limited by a wide range of particularities and idiosyncrasies.
As with all interview-based studies, our findings are shaped by the scope of our sample, participants' subjectivity and self-presentation, and the interpretative nature of a qualitative analysis that relies on ourselves as analytic instruments. While our sampling strategy aimed to capture a diverse set of communities across three different sociotechnical systems with diverse goals and at least some similar rules, our work cannot, and does not, claim to be representative or generalizable. Instead, our findings attempt to provide in-depth insights into rulemaking from the perspectives of a subset of online community leaders.

Future work should validate and extend our grounded theory using other methods.
More formal hypothesis testing would also complement the work we have done here. While we find evidence that institutional isomorphism may shape the adoption of similar rules, longitudinal studies of rule diffusion across community networks should more rigorously evaluate such a claim. Additionally, an ecological analysis of community rules could investigate whether and how rules are used to construct niches within platforms or more complex community ecosystems. %

Despite their differences, aspects of the empirical settings (Reddit, Fandom, and the Fediverse) may have shaped our findings.
Future work can extend these findings by analyzing communities on other platforms.  Although it served an important analytic purpose, it seems very possible that our focus on communities with overlapping rule sets has selected for particularities that have shaped our findings in ways that we cannot anticipate on the basis of the evidence considered here.
Additionally, idiosyncratic aspects of the three platforms at the specific point in time at which we gathered data may have determined aspects of our findings. For example, changes to Twitter/X in 2022 corresponded with a migration of many users to the Fediverse, and this may have affected rule adoption patterns and broader governance norms in that setting while we were interviewing leaders. 

Our sample of communities reflects biases towards survivorship and scale. Communities that failed or collapsed prior to the point at which we conducted our study are, by necessity, excluded from analysis. Likewise, communities that lack rules (for whatever reason) were also excluded. Consequently, our results do not provide a valid basis for inferences about whether (these particular approaches to) rules and rulemaking determine community success or survival. 

Because we only interviewed leaders (i.e., admins and moderators), future work is needed to assess how our results do or do not reflect other community members' perspectives and experiences related to rules. Simply put, community leaders may report faulty, selective, or otherwise biased or invalid representations of the processes we investigate. We chose to emphasize leaders' perspectives because of their unique role and expertise in rules and rulemaking in the communities they lead, but they may be fallible and unreliable in a variety of ways. Data sourced from community members could complement information from leaders, especially with respect to how members select communities to join, their perceptions of legitimate community governance, and how they navigate the challenges of voicing problems in online communities to inspire change.

Finally, our study uses new institutionalism and organizational ecology to conceptualize relational aspects of online communities' rulemaking. While we argue that these theories shed new light on governance in online communities, our study's scope limits our depth of engagement with these literatures and the theoretical debates therein. Future work might draw more deeply from organizational sociology to contribute new understandings of the macro-level processes shaping the development of online platforms.

\section{Conclusion}
This study set out to understand the origins of rules in online communities. We conducted a comparative, grounded theory-based analysis of 40 in-depth interviews with community leaders from subreddits, Fandom wikis, and Fediverse servers, sampled to ensure rule overlap and heterogeneity. We argue that rules and rulemaking are operational for moderating content and behavior at the same time as they also allow community leaders to establish and manage community legitimacy and identity. These findings emphasize relational dimensions of online community governance consistent with sociological explanations of institutional isomorphism and ecological dynamics. Our findings also contrast with two prevailing approaches in social computing, both of which focus on operational aspects of rules and largely overlook or take for granted the source of rules. We observed that rule creation and change practices are influenced by the sociotechnical design of the platform hosting the online community space, which may shape how problems are surfaced and who participates in problem-solving deliberation and decision-making. Rule change was often driven by problems emerging during community growth or decline, but hinged on the limited capacity of volunteer leaders, whether mechanisms for member feedback exist, and the relational dynamics between leaders and members. Together, our analysis reveals design opportunities to support effective online community governance given the relational origins of rules and rulemaking.

\begin{acks}
Special thanks to Molly DeBlanc, who participated in the early stages of this project with a focus on free/libre open source software (FLOSS) before her project split off into a separate study.
This work was also conducted with a group of undergraduate research assistants who participated in data collection and preliminary analysis and who were instrumental to the project:
Marlene Alanis, 
Paz Baum, 
Dylan Griffin, 
Amy Guo, 
Noah Hellyer, 
Aaron Phan,
Eric Rosin, 
Caroline Rygg,
Davida Yalley, 
Xiyu (Olivia) Yan,
Yiqi (Grace) Zhu, 
Carolyn Zou, 
Emily Zou. 
This work was financially supported by the US National Science Foundation, including support from the NSF Research Experiences for Undergraduates (REU) program (grants IIS-2045055, IIS-1910202, IIS-1908850, and DGE-2234667). %
\end{acks}

\bibliographystyle{ACM-Reference-Format}

\bibliography{references}


\begin{thebibliography}{102}


\ifx \showCODEN    \undefined \def \showCODEN     #1{\unskip}     \fi
\ifx \showISBNx    \undefined \def \showISBNx     #1{\unskip}     \fi
\ifx \showISBNxiii \undefined \def \showISBNxiii  #1{\unskip}     \fi
\ifx \showISSN     \undefined \def \showISSN      #1{\unskip}     \fi
\ifx \showLCCN     \undefined \def \showLCCN      #1{\unskip}     \fi
\ifx \shownote     \undefined \def \shownote      #1{#1}          \fi
\ifx \showarticletitle \undefined \def \showarticletitle #1{#1}   \fi
\ifx \showURL      \undefined \def \showURL       {\relax}        \fi
\providecommand\bibfield[2]{#2}
\providecommand\bibinfo[2]{#2}
\providecommand\natexlab[1]{#1}
\providecommand\showeprint[2][]{arXiv:#2}

\bibitem[Anaobi et~al\mbox{.}(2023)]%
        {anaobi_willadmins_2023}
\bibfield{author}{\bibinfo{person}{Ishaku~Hassan Anaobi},
  \bibinfo{person}{Aravindh Raman}, \bibinfo{person}{Ignacio Castro},
  \bibinfo{person}{Haris~Bin Zia}, \bibinfo{person}{Damilola Ibosiola}, {and}
  \bibinfo{person}{Gareth Tyson}.} \bibinfo{year}{2023}\natexlab{}.
\newblock \showarticletitle{Will {{Admins Cope}}? {{Decentralized Moderation}}
  in the {{Fediverse}}}. In \bibinfo{booktitle}{\emph{Proceedings of the {{ACM
  Web Conference}} 2023}} \emph{(\bibinfo{series}{{{WWW}} '23})}.
  \bibinfo{publisher}{Association for Computing Machinery},
  \bibinfo{address}{New York, NY, USA}, \bibinfo{pages}{3109--3120}.
\newblock
\showISBNx{978-1-4503-9416-1}
\href{https://doi.org/10.1145/3543507.3583487}{doi:\nolinkurl{10.1145/3543507.3583487}}


\bibitem[Andrade~Nunes(2023)]%
        {andradenunes_usersentiments_2023}
\bibfield{author}{\bibinfo{person}{Thatiany Andrade~Nunes}.}
  \bibinfo{year}{2023}\natexlab{}.
\newblock \showarticletitle{User {{Sentiments}} and {{Dynamics}} in the
  {{Decentralized Web}}: {{Reddit Migration}}'s {{Impact}} on {{Lemmy}}}.
\newblock \bibinfo{journal}{\emph{Journal of Multimedia Information System}}
  \bibinfo{volume}{10}, \bibinfo{number}{4} (\bibinfo{year}{2023}),
  \bibinfo{pages}{333--350}.
\newblock
\showISSN{2383-7632}
\href{https://doi.org/10.33851/JMIS.2023.10.4.333}{doi:\nolinkurl{10.33851/JMIS.2023.10.4.333}}


\bibitem[Bednar(2023)]%
        {bednar_governancehuman_2023}
\bibfield{author}{\bibinfo{person}{Jenna Bednar}.}
  \bibinfo{year}{2023}\natexlab{}.
\newblock \showarticletitle{Governance for Human Social Flourishing}.
\newblock \bibinfo{journal}{\emph{Daedalus}} \bibinfo{volume}{152},
  \bibinfo{number}{1} (\bibinfo{date}{Feb.} \bibinfo{year}{2023}),
  \bibinfo{pages}{31--45}.
\newblock
\showISSN{0011-5266}
\href{https://doi.org/10.1162/daed_a_01958}{doi:\nolinkurl{10.1162/daed_a_01958}}


\bibitem[Beschastnikh et~al\mbox{.}(2008)]%
        {beschastnikh_wikipedianselfgovernance_2008}
\bibfield{author}{\bibinfo{person}{Ivan Beschastnikh}, \bibinfo{person}{Travis
  Kriplean}, {and} \bibinfo{person}{David~W. McDonald}.}
  \bibinfo{year}{2008}\natexlab{}.
\newblock \showarticletitle{Wikipedian Self-Governance in Action: Motivating
  the Policy Lens}. In \bibinfo{booktitle}{\emph{Proceedings of the
  {{International AAAI Conference}} on {{Web}} and {{Social Media}}}},
  Vol.~\bibinfo{volume}{2}. \bibinfo{publisher}{AAAI}, \bibinfo{address}{New
  York, NY, USA}, \bibinfo{pages}{27--35}.
\newblock


\bibitem[Black et~al\mbox{.}(2008)]%
        {black_wikipedianot_2008}
\bibfield{author}{\bibinfo{person}{Laura~W. Black}, \bibinfo{person}{Ted
  Welser}, \bibinfo{person}{Jocelyn~M. DeGroot}, {and} \bibinfo{person}{D.
  Cosley}.} \bibinfo{year}{2008}\natexlab{}.
\newblock \showarticletitle{"{{Wikipedia}} Is Not a Democracy": Deliberation
  and Policy-Making in an Online Community}. In
  \bibinfo{booktitle}{\emph{Annual {{Meeting}} of the {{International
  Communication Association}}}}. \bibinfo{publisher}{ICA},
  \bibinfo{address}{Montreal, Quebec, Canada}, \bibinfo{pages}{50}.
\newblock


\bibitem[Bono et~al\mbox{.}(2024)]%
        {bono_explorationdecentralized_2024}
\bibfield{author}{\bibinfo{person}{Carlo~Alberto Bono}, \bibinfo{person}{Lucio
  La~Cava}, \bibinfo{person}{Luca Luceri}, {and} \bibinfo{person}{Francesco
  Pierri}.} \bibinfo{year}{2024}\natexlab{}.
\newblock \showarticletitle{An {{Exploration}} of {{Decentralized Moderation}}
  on {{Mastodon}}}. In \bibinfo{booktitle}{\emph{Proceedings of the 16th {{ACM
  Web Science Conference}}}} \emph{(\bibinfo{series}{{{WEBSCI}} '24})}.
  \bibinfo{publisher}{Association for Computing Machinery},
  \bibinfo{address}{New York, NY, USA}, \bibinfo{pages}{53--58}.
\newblock
\showISBNx{979-8-4007-0334-8}
\href{https://doi.org/10.1145/3614419.3644016}{doi:\nolinkurl{10.1145/3614419.3644016}}


\bibitem[Bowen(2006)]%
        {bowen_groundedtheory_2006}
\bibfield{author}{\bibinfo{person}{Glenn~A. Bowen}.}
  \bibinfo{year}{2006}\natexlab{}.
\newblock \showarticletitle{Grounded {{Theory}} and {{Sensitizing Concepts}}}.
\newblock  \bibinfo{volume}{5}, \bibinfo{number}{3} (\bibinfo{year}{2006}),
  \bibinfo{pages}{12--23}.
\newblock
\showISSN{1609-4069, 1609-4069}
\href{https://doi.org/10.1177/160940690600500304}{doi:\nolinkurl{10.1177/160940690600500304}}


\bibitem[Braun and Clarke(2006)]%
        {braun_usingthematic_2006}
\bibfield{author}{\bibinfo{person}{Virginia Braun} {and}
  \bibinfo{person}{Victoria Clarke}.} \bibinfo{year}{2006}\natexlab{}.
\newblock \showarticletitle{Using Thematic Analysis in Psychology}.
\newblock \bibinfo{journal}{\emph{Qualitative Research in Psychology}}
  \bibinfo{volume}{3}, \bibinfo{number}{2} (\bibinfo{date}{Jan.}
  \bibinfo{year}{2006}), \bibinfo{pages}{77--101}.
\newblock
\showISSN{1478-0887}
\href{https://doi.org/10.1191/1478088706qp063oa}{doi:\nolinkurl{10.1191/1478088706qp063oa}}


\bibitem[Butler et~al\mbox{.}(2008)]%
        {butler_dontlook_2008}
\bibfield{author}{\bibinfo{person}{Brian~S. Butler}, \bibinfo{person}{Elisabeth
  Joyce}, {and} \bibinfo{person}{Jacqueline Pike}.}
  \bibinfo{year}{2008}\natexlab{}.
\newblock \showarticletitle{Don't Look Now, but We've Created a Bureaucracy:
  The Nature and Roles of Policies and Rules in {{Wikipedia}}}. In
  \bibinfo{booktitle}{\emph{Proceedings of the {{SIGCHI Conference}} on {{Human
  Factors}} in {{Computing Systems}}}} \emph{(\bibinfo{series}{{{CHI}} '08})}.
  \bibinfo{publisher}{ACM}, \bibinfo{address}{New York, NY, USA},
  \bibinfo{pages}{1101--1110}.
\newblock
\showISBNx{978-1-60558-011-1}
\href{https://doi.org/10.1145/1357054.1357227}{doi:\nolinkurl{10.1145/1357054.1357227}}


\bibitem[Caelin(2022)]%
        {caelin_decentralizednetworks_2022}
\bibfield{author}{\bibinfo{person}{Derek Caelin}.}
  \bibinfo{year}{2022}\natexlab{}.
\newblock \showarticletitle{Decentralized {{Networks}} vs {{The Trolls}}}.
\newblock In \bibinfo{booktitle}{\emph{Fundamental {{Challenges}} to {{Global
  Peace}} and {{Security}} : {{The Future}} of {{Humanity}}}},
  \bibfield{editor}{\bibinfo{person}{Hoda Mahmoudi},
  \bibinfo{person}{Michael~H. Allen}, {and} \bibinfo{person}{Kate Seaman}}
  (Eds.). \bibinfo{publisher}{Springer International Publishing},
  \bibinfo{address}{Cham}, \bibinfo{pages}{143--168}.
\newblock
\showISBNx{978-3-030-79072-1}
\href{https://doi.org/10.1007/978-3-030-79072-1_8}{doi:\nolinkurl{10.1007/978-3-030-79072-1_8}}


\bibitem[Caplan and {boyd}(2018)]%
        {caplan_isomorphismalgorithms_2018}
\bibfield{author}{\bibinfo{person}{Robyn Caplan} {and} \bibinfo{person}{danah
  {boyd}}.} \bibinfo{year}{2018}\natexlab{}.
\newblock \showarticletitle{Isomorphism through Algorithms: {{Institutional}}
  Dependencies in the Case of {{Facebook}}}.
\newblock \bibinfo{journal}{\emph{Big Data \& Society}} \bibinfo{volume}{5},
  \bibinfo{number}{1} (\bibinfo{date}{Jan.} \bibinfo{year}{2018}),
  \bibinfo{pages}{12}.
\newblock
\showISSN{2053-9517}
\href{https://doi.org/10.1177/2053951718757253}{doi:\nolinkurl{10.1177/2053951718757253}}


\bibitem[Carr and Hayes(2015)]%
        {carr_socialmedia_2015}
\bibfield{author}{\bibinfo{person}{Caleb~T. Carr} {and}
  \bibinfo{person}{Rebecca~A. Hayes}.} \bibinfo{year}{2015}\natexlab{}.
\newblock \showarticletitle{Social {{Media}}: {{Defining}}, {{Developing}}, and
  {{Divining}}}.
\newblock \bibinfo{journal}{\emph{Atlantic Journal of Communication}}
  \bibinfo{volume}{23}, \bibinfo{number}{1} (\bibinfo{date}{Jan.}
  \bibinfo{year}{2015}), \bibinfo{pages}{46--65}.
\newblock
\showISSN{1545-6870, 1545-6889}
\href{https://doi.org/10.1080/15456870.2015.972282}{doi:\nolinkurl{10.1080/15456870.2015.972282}}


\bibitem[Carroll(1985)]%
        {carroll_concentrationspecialization_1985}
\bibfield{author}{\bibinfo{person}{Glenn~R. Carroll}.}
  \bibinfo{year}{1985}\natexlab{}.
\newblock \showarticletitle{Concentration and Specialization: {{Dynamics}} of
  Niche Width in Populations of Organizations}.
\newblock \bibinfo{journal}{\emph{Amer. J. Sociology}} \bibinfo{volume}{90},
  \bibinfo{number}{6} (\bibinfo{date}{May} \bibinfo{year}{1985}),
  \bibinfo{pages}{1262--1283}.
\newblock
\showISSN{0002-9602}
\href{https://doi.org/10.1086/228210}{doi:\nolinkurl{10.1086/228210}}


\bibitem[Carroll and Swaminathan(2000)]%
        {carroll_whymicrobrewery_2000}
\bibfield{author}{\bibinfo{person}{Glenn~R. Carroll} {and}
  \bibinfo{person}{Anand Swaminathan}.} \bibinfo{year}{2000}\natexlab{}.
\newblock \showarticletitle{Why the Microbrewery Movement? {{Organizational}}
  Dynamics of Resource Partitioning in the {{U}}.{{S}}. Brewing Industry}.
\newblock \bibinfo{journal}{\emph{Amer. J. Sociology}} \bibinfo{volume}{106},
  \bibinfo{number}{3} (\bibinfo{year}{2000}), \bibinfo{pages}{715--762}.
\newblock
\showISSN{0002-9602}
\href{https://doi.org/10.1086/318962}{doi:\nolinkurl{10.1086/318962}}
\showeprint[jstor]{10.1086/318962}


\bibitem[Cava et~al\mbox{.}(2023)]%
        {cava_driverssocial_2023}
\bibfield{author}{\bibinfo{person}{Lucio~La Cava}, \bibinfo{person}{Luca~Maria
  Aiello}, {and} \bibinfo{person}{Andrea Tagarelli}.}
  \bibinfo{year}{2023}\natexlab{}.
\newblock \showarticletitle{Drivers of Social Influence in the {{Twitter}}
  Migration to {{Mastodon}}}.
\newblock \bibinfo{journal}{\emph{Scientific Reports}} \bibinfo{volume}{13},
  \bibinfo{number}{1} (\bibinfo{date}{Dec.} \bibinfo{year}{2023}),
  \bibinfo{pages}{21626}.
\newblock
\showISSN{2045-2322}
\href{https://doi.org/10.1038/s41598-023-48200-7}{doi:\nolinkurl{10.1038/s41598-023-48200-7}}


\bibitem[Chandrasekharan et~al\mbox{.}(2019)]%
        {chandrasekharan_crossmodcrosscommunity_2019}
\bibfield{author}{\bibinfo{person}{Eshwar Chandrasekharan},
  \bibinfo{person}{Chaitrali Gandhi}, \bibinfo{person}{Matthew~Wortley
  Mustelier}, {and} \bibinfo{person}{Eric Gilbert}.}
  \bibinfo{year}{2019}\natexlab{}.
\newblock \showarticletitle{Crossmod: A Cross-Community Learning-Based System
  to Assist Reddit Moderators}.
\newblock \bibinfo{journal}{\emph{Proceedings of the ACM on Human-Computer
  Interaction}} \bibinfo{volume}{3}, \bibinfo{number}{CSCW}
  (\bibinfo{date}{Nov.} \bibinfo{year}{2019}), \bibinfo{pages}{1--30}.
\newblock
\showISSN{2573-0142}
\href{https://doi.org/10.1145/3359276}{doi:\nolinkurl{10.1145/3359276}}


\bibitem[Chandrasekharan et~al\mbox{.}(2017)]%
        {chandrasekharan_youcant_2017}
\bibfield{author}{\bibinfo{person}{Eshwar Chandrasekharan},
  \bibinfo{person}{Umashanthi Pavalanathan}, \bibinfo{person}{Anirudh
  Srinivasan}, \bibinfo{person}{Adam Glynn}, \bibinfo{person}{Jacob
  Eisenstein}, {and} \bibinfo{person}{Eric Gilbert}.}
  \bibinfo{year}{2017}\natexlab{}.
\newblock \showarticletitle{You Can't Stay Here: {{The}} Efficacy of Reddit's
  2015 Ban Examined through Hate Speech}.
\newblock \bibinfo{journal}{\emph{Proc. ACM Hum.-Comput. Interact.}}
  \bibinfo{volume}{1}, \bibinfo{number}{CSCW} (\bibinfo{date}{Dec.}
  \bibinfo{year}{2017}), \bibinfo{pages}{31:1--31:22}.
\newblock
\showISSN{2573-0142}
\href{https://doi.org/10.1145/3134666}{doi:\nolinkurl{10.1145/3134666}}


\bibitem[Chandrasekharan et~al\mbox{.}(2018)]%
        {chandrasekharan_internetshidden_2018}
\bibfield{author}{\bibinfo{person}{Eshwar Chandrasekharan},
  \bibinfo{person}{Mattia Samory}, \bibinfo{person}{Shagun Jhaver},
  \bibinfo{person}{Hunter Charvat}, \bibinfo{person}{Amy Bruckman},
  \bibinfo{person}{Cliff Lampe}, \bibinfo{person}{Jacob Eisenstein}, {and}
  \bibinfo{person}{Eric Gilbert}.} \bibinfo{year}{2018}\natexlab{}.
\newblock \showarticletitle{The {{Internet}}'s Hidden Rules: {{An}} Empirical
  Study of {{Reddit}} Norm Violations at Micro, Meso, and Macro Scales}.
\newblock \bibinfo{journal}{\emph{Proc. ACM Hum.-Comput. Interact.}}
  \bibinfo{volume}{2}, \bibinfo{number}{CSCW} (\bibinfo{year}{2018}),
  \bibinfo{pages}{32:1--32:25}.
\newblock
\showISSN{2573-0142}
\href{https://doi.org/10.1145/3274301}{doi:\nolinkurl{10.1145/3274301}}


\bibitem[Charmaz(2015)]%
        {charmaz_constructinggrounded_2015}
\bibfield{author}{\bibinfo{person}{Kathy Charmaz}.}
  \bibinfo{year}{2015}\natexlab{}.
\newblock \bibinfo{booktitle}{\emph{Constructing Grounded Theory: {{A}}
  Practical Guide through Qualitative Analysis} (\bibinfo{edition}{2nd} ed.)}.
\newblock \bibinfo{publisher}{SAGE}, \bibinfo{address}{Thousand Oaks,
  California}.
\newblock
\showISBNx{0-7619-7352-4}
\showLCCN{H61.24.C45 2006}


\bibitem[Colglazier et~al\mbox{.}(2024)]%
        {colglazier_effectsgroup_2024}
\bibfield{author}{\bibinfo{person}{Carl Colglazier}, \bibinfo{person}{Nathan
  TeBlunthuis}, {and} \bibinfo{person}{Aaron Shaw}.}
  \bibinfo{year}{2024}\natexlab{}.
\newblock \showarticletitle{The Effects of Group Sanctions on Participation and
  Toxicity: {{Quasi-experimental}} Evidence from the Fediverse}.
\newblock \bibinfo{journal}{\emph{Proceedings of the International AAAI
  Conference on Web and Social Media}}  \bibinfo{volume}{18}
  (\bibinfo{date}{May} \bibinfo{year}{2024}), \bibinfo{pages}{315--328}.
\newblock
\showISSN{2334-0770}
\href{https://doi.org/10.1609/icwsm.v18i1.31316}{doi:\nolinkurl{10.1609/icwsm.v18i1.31316}}


\bibitem[Crawford and Gillespie(2016)]%
        {crawford_whatflag_2016}
\bibfield{author}{\bibinfo{person}{Kate Crawford} {and}
  \bibinfo{person}{Tarleton Gillespie}.} \bibinfo{year}{2016}\natexlab{}.
\newblock \showarticletitle{What Is a Flag for? {{Social}} Media Reporting
  Tools and the Vocabulary of Complaint}.
\newblock \bibinfo{journal}{\emph{New Media \& Society}} \bibinfo{volume}{18},
  \bibinfo{number}{3} (\bibinfo{date}{March} \bibinfo{year}{2016}),
  \bibinfo{pages}{410--428}.
\newblock
\showISSN{1461-4448, 1461-7315}
\href{https://doi.org/10.1177/1461444814543163}{doi:\nolinkurl{10.1177/1461444814543163}}


\bibitem[Cullen and Kairam(2022)]%
        {cullen_practicingmoderation_2022}
\bibfield{author}{\bibinfo{person}{Amanda L.~L. Cullen} {and}
  \bibinfo{person}{Sanjay~R. Kairam}.} \bibinfo{year}{2022}\natexlab{}.
\newblock \showarticletitle{Practicing {{Moderation}}: {{Community Moderation}}
  as {{Reflective Practice}}}.
\newblock \bibinfo{journal}{\emph{Proc. ACM Hum.-Comput. Interact.}}
  \bibinfo{volume}{6}, \bibinfo{number}{CSCW1} (\bibinfo{date}{April}
  \bibinfo{year}{2022}), \bibinfo{pages}{111:1--111:32}.
\newblock
\href{https://doi.org/10.1145/3512958}{doi:\nolinkurl{10.1145/3512958}}


\bibitem[Cyert and March(1963)]%
        {cyert_behavioraltheory_1963}
\bibfield{author}{\bibinfo{person}{Richard~Michael Cyert} {and}
  \bibinfo{person}{James~G. March}.} \bibinfo{year}{1963}\natexlab{}.
\newblock \bibinfo{booktitle}{\emph{A Behavioral Theory of the Firm}}.
\newblock \bibinfo{publisher}{Prentice-Hall}, \bibinfo{address}{Englewood
  Cliffs, N.J.}
\newblock
\showISBNx{978-0-13-073304-7}
\showLCCN{HD38 C9, HD30.23 .C9, HD38.C9, HD38 .C9, HD38 .C9 1963}


\bibitem[{de Laat}(2012)]%
        {delaat_coercionempowerment_2012}
\bibfield{author}{\bibinfo{person}{Paul~B. {de Laat}}.}
  \bibinfo{year}{2012}\natexlab{}.
\newblock \showarticletitle{Coercion or Empowerment? {{Moderation}} of Content
  in {{Wikipedia}} as `Essentially Contested' Bureaucratic Rules}.
\newblock \bibinfo{journal}{\emph{Ethics and Information Technology}}
  \bibinfo{volume}{14}, \bibinfo{number}{2} (\bibinfo{date}{June}
  \bibinfo{year}{2012}), \bibinfo{pages}{123--135}.
\newblock
\showISSN{1572-8439}
\href{https://doi.org/10.1007/s10676-012-9289-7}{doi:\nolinkurl{10.1007/s10676-012-9289-7}}


\bibitem[DiMaggio and Powell(1983)]%
        {dimaggio_ironcage_1983}
\bibfield{author}{\bibinfo{person}{Paul~J. DiMaggio} {and}
  \bibinfo{person}{Walter~W. Powell}.} \bibinfo{year}{1983}\natexlab{}.
\newblock \showarticletitle{The {{Iron Cage Revisited}}: {{Institutional
  Isomorphism}} and {{Collective Rationality}} in {{Organizational Fields}}}.
\newblock \bibinfo{journal}{\emph{American Sociological Review}}
  \bibinfo{volume}{48}, \bibinfo{number}{2} (\bibinfo{year}{1983}),
  \bibinfo{pages}{147--160}.
\newblock
\showISSN{0003-1224}
\href{https://doi.org/10.2307/2095101}{doi:\nolinkurl{10.2307/2095101}}
\showeprint[jstor]{2095101}


\bibitem[Dosono and Semaan(2019)]%
        {dosono_moderationpractices_2019}
\bibfield{author}{\bibinfo{person}{Bryan Dosono} {and} \bibinfo{person}{Bryan
  Semaan}.} \bibinfo{year}{2019}\natexlab{}.
\newblock \showarticletitle{Moderation Practices as Emotional Labor in
  Sustaining Online Communities: The Case of Aapi Identity Work on Reddit}. In
  \bibinfo{booktitle}{\emph{Proceedings of the 2019 {{CHI Conference}} on
  {{Human Factors}} in {{Computing Systems}}}} \emph{(\bibinfo{series}{{{CHI}}
  '19})}. \bibinfo{publisher}{Association for Computing Machinery},
  \bibinfo{address}{Glasgow, Scotland Uk}, \bibinfo{pages}{1--13}.
\newblock
\showISBNx{978-1-4503-5970-2}
\href{https://doi.org/10.1145/3290605.3300372}{doi:\nolinkurl{10.1145/3290605.3300372}}


\bibitem[{Dunbar-Hester}(2024)]%
        {dunbar-hester_showingyour_2024}
\bibfield{author}{\bibinfo{person}{Christina {Dunbar-Hester}}.}
  \bibinfo{year}{2024}\natexlab{}.
\newblock \showarticletitle{Showing Your Ass on {{Mastodon}}: {{Lossy}}
  Distribution, Hashtag Activism, and Public Scrutiny on Federated, Feral
  Social Media}.
\newblock \bibinfo{journal}{\emph{First Monday}} \bibinfo{volume}{29},
  \bibinfo{number}{3 - 4} (\bibinfo{date}{March} \bibinfo{year}{2024}),
  \bibinfo{pages}{29}.
\newblock
\showISSN{1396-0466}
\href{https://doi.org/10.5210/fm.v29i3.13367}{doi:\nolinkurl{10.5210/fm.v29i3.13367}}


\bibitem[Fiesler et~al\mbox{.}(2018)]%
        {fiesler_redditrules_2018}
\bibfield{author}{\bibinfo{person}{Casey Fiesler},
  \bibinfo{person}{Jialun"~Aaron" Jiang}, \bibinfo{person}{Joshua McCann},
  \bibinfo{person}{Kyle Frye}, {and} \bibinfo{person}{Jed~R. Brubaker}.}
  \bibinfo{year}{2018}\natexlab{}.
\newblock \showarticletitle{Reddit Rules! {{Characterizing}} an Ecosystem of
  Governance.}. In \bibinfo{booktitle}{\emph{Proceedings of the {{International
  AAAI Conference}} on {{Web}} and {{Social Media}}}},
  Vol.~\bibinfo{volume}{12}. \bibinfo{publisher}{AAAI},
  \bibinfo{address}{Stanford, CA}, \bibinfo{pages}{72--81}.
\newblock
\href{https://doi.org/10.1609/icwsm.v12i1.15033}{doi:\nolinkurl{10.1609/icwsm.v12i1.15033}}


\bibitem[Frey et~al\mbox{.}(2019)]%
        {frey_thisplace_2019}
\bibfield{author}{\bibinfo{person}{Seth Frey}, \bibinfo{person}{P.~M. Krafft},
  {and} \bibinfo{person}{Brian~C. Keegan}.} \bibinfo{year}{2019}\natexlab{}.
\newblock \showarticletitle{"{{This Place Does What It Was Built For}}":
  {{Designing Digital Institutions}} for {{Participatory Change}}}.
\newblock \bibinfo{journal}{\emph{Proc. ACM Hum.-Comput. Interact.}}
  \bibinfo{volume}{3}, \bibinfo{number}{CSCW} (\bibinfo{date}{Nov.}
  \bibinfo{year}{2019}), \bibinfo{pages}{32:1--32:31}.
\newblock
\href{https://doi.org/10.1145/3359134}{doi:\nolinkurl{10.1145/3359134}}


\bibitem[Frey and Schneider(2021)]%
        {frey_effectivevoice_2021}
\bibfield{author}{\bibinfo{person}{Seth Frey} {and} \bibinfo{person}{Nathan
  Schneider}.} \bibinfo{year}{2021}\natexlab{}.
\newblock \showarticletitle{Effective Voice: {{Beyond}} Exit and Affect in
  Online Communities}.
\newblock \bibinfo{journal}{\emph{New Media \& Society}} \bibinfo{volume}{25},
  \bibinfo{number}{9} (\bibinfo{date}{Sept.} \bibinfo{year}{2021}),
  \bibinfo{pages}{14614448211044025}.
\newblock
\showISSN{1461-4448}
\href{https://doi.org/10.1177/14614448211044025}{doi:\nolinkurl{10.1177/14614448211044025}}


\bibitem[Frey and Sumner(2019)]%
        {frey_emergenceintegrated_2019}
\bibfield{author}{\bibinfo{person}{Seth Frey} {and} \bibinfo{person}{Robert~W.
  Sumner}.} \bibinfo{year}{2019}\natexlab{}.
\newblock \showarticletitle{Emergence of Integrated Institutions in a Large
  Population of Self-Governing Communities}.
\newblock \bibinfo{journal}{\emph{PLOS ONE}} \bibinfo{volume}{14},
  \bibinfo{number}{7} (\bibinfo{date}{July} \bibinfo{year}{2019}),
  \bibinfo{pages}{e0216335}.
\newblock
\showISSN{1932-6203}
\href{https://doi.org/10.1371/journal.pone.0216335}{doi:\nolinkurl{10.1371/journal.pone.0216335}}


\bibitem[Gehl and Zulli(2023)]%
        {gehl_digitalcovenant_2023}
\bibfield{author}{\bibinfo{person}{Robert~W. Gehl} {and} \bibinfo{person}{Diana
  Zulli}.} \bibinfo{year}{2023}\natexlab{}.
\newblock \showarticletitle{The Digital Covenant: Non-Centralized Platform
  Governance on the Mastodon Social Network}.
\newblock \bibinfo{journal}{\emph{Information, Communication \& Society}}
  \bibinfo{volume}{26}, \bibinfo{number}{16} (\bibinfo{date}{Dec.}
  \bibinfo{year}{2023}), \bibinfo{pages}{3275--3291}.
\newblock
\showISSN{1369-118X}
\href{https://doi.org/10.1080/1369118X.2022.2147400}{doi:\nolinkurl{10.1080/1369118X.2022.2147400}}


\bibitem[Geiger(2014)]%
        {geiger_botsbespoke_2014}
\bibfield{author}{\bibinfo{person}{R.~Stuart Geiger}.}
  \bibinfo{year}{2014}\natexlab{}.
\newblock \showarticletitle{Bots, Bespoke, Code and the Materiality of Software
  Platforms}.
\newblock \bibinfo{journal}{\emph{Information, Communication \& Society}}
  \bibinfo{volume}{17}, \bibinfo{number}{3} (\bibinfo{date}{March}
  \bibinfo{year}{2014}), \bibinfo{pages}{342--356}.
\newblock
\showISSN{1369-118X}
\href{https://doi.org/10.1080/1369118X.2013.873069}{doi:\nolinkurl{10.1080/1369118X.2013.873069}}


\bibitem[Giles et~al\mbox{.}(2013)]%
        {giles_timing_2013}
\bibfield{author}{\bibinfo{person}{Tracey Giles}, \bibinfo{person}{Lindy King},
  {and} \bibinfo{person}{Sheryl {de Lacey}}.} \bibinfo{year}{April/June
  2013}\natexlab{}.
\newblock \showarticletitle{The {{Timing}} of the {{Literature Review}} in
  {{Grounded Theory Research}}: {{An Open Mind Versus}} an {{Empty Head}}}.
\newblock \bibinfo{journal}{\emph{Advances in Nursing Science}}
  \bibinfo{volume}{36}, \bibinfo{number}{2} (\bibinfo{year}{April/June 2013}),
  \bibinfo{pages}{E29}.
\newblock
\showISSN{0161-9268}
\href{https://doi.org/10.1097/ANS.0b013e3182902035}{doi:\nolinkurl{10.1097/ANS.0b013e3182902035}}


\bibitem[Gillespie(2018)]%
        {gillespie_custodiansinternet_2018}
\bibfield{author}{\bibinfo{person}{Tarleton Gillespie}.}
  \bibinfo{year}{2018}\natexlab{}.
\newblock \bibinfo{booktitle}{\emph{Custodians of the Internet: Platforms,
  Content Moderation, and the Hidden Decisions That Shape Social Media}}.
\newblock \bibinfo{publisher}{Yale University Press}, \bibinfo{address}{New
  Haven}.
\newblock
\showISBNx{978-0-300-17313-0}
\showLCCN{HM742 .G575 2018}


\bibitem[Grimmelmann(2017)]%
        {grimmelmann_virtuesmoderation_2017}
\bibfield{author}{\bibinfo{person}{James Grimmelmann}.}
  \bibinfo{year}{2017}\natexlab{}.
\newblock \bibinfo{booktitle}{\emph{The {{Virtues}} of {{Moderation}}}}.
\newblock \bibinfo{type}{Preprint}. \bibinfo{institution}{LawArXiv}.
\newblock
\href{https://doi.org/10.31228/osf.io/qwxf5}{doi:\nolinkurl{10.31228/osf.io/qwxf5}}


\bibitem[Halfaker and Geiger(2020)]%
        {halfaker_oreslowering_2020}
\bibfield{author}{\bibinfo{person}{Aaron Halfaker} {and}
  \bibinfo{person}{R.~Stuart Geiger}.} \bibinfo{year}{2020}\natexlab{}.
\newblock \showarticletitle{{{ORES}}: {{Lowering Barriers}} with
  {{Participatory Machine Learning}} in {{Wikipedia}}}.
\newblock \bibinfo{journal}{\emph{Proceedings of the ACM on Human-Computer
  Interaction}} \bibinfo{volume}{4}, \bibinfo{number}{CSCW2}
  (\bibinfo{year}{2020}), \bibinfo{pages}{37}.
\newblock
\href{https://doi.org/10.1145/3415219}{doi:\nolinkurl{10.1145/3415219}}


\bibitem[Halfaker et~al\mbox{.}(2013)]%
        {halfaker_risedecline_2013}
\bibfield{author}{\bibinfo{person}{Aaron Halfaker}, \bibinfo{person}{R.~Stuart
  Geiger}, \bibinfo{person}{Jonathan~T. Morgan}, {and} \bibinfo{person}{John
  Riedl}.} \bibinfo{year}{2013}\natexlab{}.
\newblock \showarticletitle{The Rise and Decline of an Open Collaboration
  System: How {{Wikipedia}}'s Reaction to Popularity Is Causing Its Decline}.
\newblock \bibinfo{journal}{\emph{American Behavioral Scientist}}
  \bibinfo{volume}{57}, \bibinfo{number}{5} (\bibinfo{date}{May}
  \bibinfo{year}{2013}), \bibinfo{pages}{664--688}.
\newblock
\showISSN{0002-7642}
\href{https://doi.org/10.1177/0002764212469365}{doi:\nolinkurl{10.1177/0002764212469365}}


\bibitem[Hannan and Freeman(1977)]%
        {hannan_populationecology_1977}
\bibfield{author}{\bibinfo{person}{Michael~T. Hannan} {and}
  \bibinfo{person}{John Freeman}.} \bibinfo{year}{1977}\natexlab{}.
\newblock \showarticletitle{The Population Ecology of Organizations}.
\newblock \bibinfo{journal}{\emph{Amer. J. Sociology}} \bibinfo{volume}{82},
  \bibinfo{number}{5} (\bibinfo{year}{1977}), \bibinfo{pages}{929--964}.
\newblock
\showISSN{0002-9602}
\href{https://doi.org/10.2307/2777807}{doi:\nolinkurl{10.2307/2777807}}
\showeprint[jstor]{2777807}


\bibitem[Hannan and Freeman(1984)]%
        {hannan_structuralinertia_1984}
\bibfield{author}{\bibinfo{person}{Michael~T. Hannan} {and}
  \bibinfo{person}{John Freeman}.} \bibinfo{year}{1984}\natexlab{}.
\newblock \showarticletitle{Structural Inertia and Organizational Change}.
\newblock \bibinfo{journal}{\emph{American Sociological Review}}
  \bibinfo{volume}{49}, \bibinfo{number}{2} (\bibinfo{date}{April}
  \bibinfo{year}{1984}), \bibinfo{pages}{149}.
\newblock
\showISSN{00031224}
\href{https://doi.org/10.2307/2095567}{doi:\nolinkurl{10.2307/2095567}}
\showeprint[jstor]{2095567}


\bibitem[He et~al\mbox{.}(2023)]%
        {he_flockingmastodon_2023}
\bibfield{author}{\bibinfo{person}{Jiahui He}, \bibinfo{person}{Haris~Bin Zia},
  \bibinfo{person}{Ignacio Castro}, \bibinfo{person}{Aravindh Raman},
  \bibinfo{person}{Nishanth Sastry}, {and} \bibinfo{person}{Gareth Tyson}.}
  \bibinfo{year}{2023}\natexlab{}.
\newblock \showarticletitle{Flocking to {{Mastodon}}: {{Tracking}} the {{Great
  Twitter Migration}}}. In \bibinfo{booktitle}{\emph{Proceedings of the 2023
  {{ACM}} on {{Internet Measurement Conference}}}}. \bibinfo{publisher}{ACM},
  \bibinfo{address}{Montreal QC Canada}, \bibinfo{pages}{111--123}.
\newblock
\showISBNx{979-8-4007-0382-9}
\href{https://doi.org/10.1145/3618257.3624819}{doi:\nolinkurl{10.1145/3618257.3624819}}


\bibitem[Hecht and Gergle(2010)]%
        {hecht_towerbabel_2010}
\bibfield{author}{\bibinfo{person}{Brent Hecht} {and} \bibinfo{person}{Darren
  Gergle}.} \bibinfo{year}{2010}\natexlab{}.
\newblock \showarticletitle{The {{Tower}} of {{Babel}} Meets {{Web}} 2.0:
  User-Generated Content and Its Applications in a Multilingual Context}. In
  \bibinfo{booktitle}{\emph{Proceedings of the {{SIGCHI}} Conference on Human
  Factors in Computing Systems}} \emph{(\bibinfo{series}{{{CHI}} '10})}.
  \bibinfo{publisher}{ACM}, \bibinfo{address}{Atlanta, Georgia, USA},
  \bibinfo{pages}{291--300}.
\newblock
\showISBNx{978-1-60558-929-9}
\href{https://doi.org/10.1145/1753326.1753370}{doi:\nolinkurl{10.1145/1753326.1753370}}


\bibitem[Horta~Ribeiro et~al\mbox{.}(2021)]%
        {hortaribeiro_platformmigrations_2021}
\bibfield{author}{\bibinfo{person}{Manoel~Horta Horta~Ribeiro},
  \bibinfo{person}{Shagun Jhaver}, \bibinfo{person}{Savvas Zannettou},
  \bibinfo{person}{Jeremy Blackburn}, \bibinfo{person}{Gianluca Stringhini},
  \bibinfo{person}{Emiliano De~Cristofaro}, {and} \bibinfo{person}{Robert
  West}.} \bibinfo{year}{2021}\natexlab{}.
\newblock \showarticletitle{Do Platform Migrations Compromise Content
  Moderation? {{Evidence}} from r/The\_donald and r/Incels}.
\newblock \bibinfo{journal}{\emph{Proceedings of the ACM on Human-Computer
  Interaction}} \bibinfo{volume}{5}, \bibinfo{number}{CSCW2}
  (\bibinfo{date}{Oct.} \bibinfo{year}{2021}), \bibinfo{pages}{316:1--316:24}.
\newblock
\href{https://doi.org/10.1145/3476057}{doi:\nolinkurl{10.1145/3476057}}


\bibitem[Hwang and Foote(2021)]%
        {hwang_whypeople_2021}
\bibfield{author}{\bibinfo{person}{Sohyeon Hwang} {and}
  \bibinfo{person}{Jeremy~D. Foote}.} \bibinfo{year}{2021}\natexlab{}.
\newblock \showarticletitle{Why Do People Participate in Small Online
  Communities?}
\newblock \bibinfo{journal}{\emph{Proceedings of the ACM on Human-Computer
  Interaction}} \bibinfo{volume}{5}, \bibinfo{number}{CSCW2}
  (\bibinfo{date}{Oct.} \bibinfo{year}{2021}), \bibinfo{pages}{1--25}.
\newblock
\showISSN{2573-0142}
\href{https://doi.org/10.1145/3479606}{doi:\nolinkurl{10.1145/3479606}}


\bibitem[Hwang et~al\mbox{.}(2024)]%
        {hwang_adoptingthirdparty_2024}
\bibfield{author}{\bibinfo{person}{Sohyeon Hwang}, \bibinfo{person}{Charles
  Kiene}, \bibinfo{person}{Serene Ong}, {and} \bibinfo{person}{Aaron Shaw}.}
  \bibinfo{year}{2024}\natexlab{}.
\newblock \showarticletitle{Adopting Third-Party Bots for Managing Online
  Communities}.
\newblock \bibinfo{journal}{\emph{Proceedings of the ACM on Human-Computer
  Interaction: Computer Supported Cooperative Work}} \bibinfo{volume}{8},
  \bibinfo{number}{CSCW1} (\bibinfo{date}{April} \bibinfo{year}{2024}),
  \bibinfo{pages}{216:1--216:26}.
\newblock
\href{https://doi.org/10.1145/3653707}{doi:\nolinkurl{10.1145/3653707}}


\bibitem[Hwang et~al\mbox{.}(2025)]%
        {hwang_trustfriction_2025}
\bibfield{author}{\bibinfo{person}{Sohyeon Hwang}, \bibinfo{person}{Priyanka
  Nanayakkara}, {and} \bibinfo{person}{Yan Shvartzshnaider}.}
  \bibinfo{year}{2025}\natexlab{}.
\newblock \bibinfo{title}{Trust and {{Friction}}: {{Negotiating How Information
  Flows Through Decentralized Social Media}}}.
\newblock
\href{https://doi.org/10.48550/arXiv.2503.02150}{doi:\nolinkurl{10.48550/arXiv.2503.02150}}
\showeprint[arxiv]{2503.02150}~[cs]


\bibitem[Hwang and Shaw(2022)]%
        {hwang_rulesrulemaking_2022}
\bibfield{author}{\bibinfo{person}{Sohyeon Hwang} {and} \bibinfo{person}{Aaron
  Shaw}.} \bibinfo{year}{2022}\natexlab{}.
\newblock \showarticletitle{Rules and Rule-Making in the Five Largest
  {{Wikipedias}}}.
\newblock \bibinfo{journal}{\emph{Proceedings of the International AAAI
  Conference on Web and Social Media}}  \bibinfo{volume}{16}
  (\bibinfo{date}{May} \bibinfo{year}{2022}), \bibinfo{pages}{347--357}.
\newblock
\showISSN{2334-0770}
\href{https://doi.org/10.1609/icwsm.v16i1.19297}{doi:\nolinkurl{10.1609/icwsm.v16i1.19297}}


\bibitem[Jeong et~al\mbox{.}(2024)]%
        {jeong_exploringplatform_2024}
\bibfield{author}{\bibinfo{person}{Ujun Jeong}, \bibinfo{person}{Paras Sheth},
  \bibinfo{person}{Anique Tahir}, \bibinfo{person}{Faisal Alatawi},
  \bibinfo{person}{H.~Russell Bernard}, {and} \bibinfo{person}{Huan Liu}.}
  \bibinfo{year}{2024}\natexlab{}.
\newblock \showarticletitle{Exploring {{Platform Migration Patterns}} between
  {{Twitter}} and {{Mastodon}}: {{A User Behavior Study}}}.
\newblock \bibinfo{journal}{\emph{Proceedings of the International AAAI
  Conference on Web and Social Media}}  \bibinfo{volume}{18}
  (\bibinfo{date}{May} \bibinfo{year}{2024}), \bibinfo{pages}{738--750}.
\newblock
\showISSN{2334-0770}
\href{https://doi.org/10.1609/icwsm.v18i1.31348}{doi:\nolinkurl{10.1609/icwsm.v18i1.31348}}


\bibitem[Jhaver et~al\mbox{.}(2019a)]%
        {jhaver_didyou_2019}
\bibfield{author}{\bibinfo{person}{Shagun Jhaver},
  \bibinfo{person}{Darren~Scott Appling}, \bibinfo{person}{Eric Gilbert}, {and}
  \bibinfo{person}{Amy Bruckman}.} \bibinfo{year}{2019}\natexlab{a}.
\newblock \showarticletitle{"{{Did}} You Suspect the Post Would Be Removed?":
  {{Understanding}} User Reactions to Content Removals on Reddit}.
\newblock \bibinfo{journal}{\emph{Proceedings of the ACM on Human-Computer
  Interaction}} \bibinfo{volume}{3}, \bibinfo{number}{CSCW}
  (\bibinfo{date}{Nov.} \bibinfo{year}{2019}), \bibinfo{pages}{192:1--192:33}.
\newblock
\href{https://doi.org/10.1145/3359294}{doi:\nolinkurl{10.1145/3359294}}


\bibitem[Jhaver et~al\mbox{.}(2019b)]%
        {jhaver_humanmachinecollaboration_2019}
\bibfield{author}{\bibinfo{person}{Shagun Jhaver}, \bibinfo{person}{Iris
  Birman}, \bibinfo{person}{Eric Gilbert}, {and} \bibinfo{person}{Amy
  Bruckman}.} \bibinfo{year}{2019}\natexlab{b}.
\newblock \showarticletitle{Human-Machine Collaboration for Content Regulation:
  The Case of Reddit Automoderator}.
\newblock \bibinfo{journal}{\emph{ACM Trans. Comput.-Hum. Interact.}}
  \bibinfo{volume}{26}, \bibinfo{number}{5} (\bibinfo{date}{July}
  \bibinfo{year}{2019}), \bibinfo{pages}{31:1--31:35}.
\newblock
\showISSN{1073-0516}
\href{https://doi.org/10.1145/3338243}{doi:\nolinkurl{10.1145/3338243}}


\bibitem[Jhaver et~al\mbox{.}(2019c)]%
        {jhaver_doestransparency_2019}
\bibfield{author}{\bibinfo{person}{Shagun Jhaver}, \bibinfo{person}{Amy
  Bruckman}, {and} \bibinfo{person}{Eric Gilbert}.}
  \bibinfo{year}{2019}\natexlab{c}.
\newblock \showarticletitle{Does Transparency in Moderation Really Matter?
  {{User}} Behavior after Content Removal Explanations on Reddit}.
\newblock \bibinfo{journal}{\emph{Proceedings of the ACM on Human-Computer
  Interaction}} \bibinfo{volume}{3}, \bibinfo{number}{CSCW}
  (\bibinfo{date}{Nov.} \bibinfo{year}{2019}), \bibinfo{pages}{150:1--150:27}.
\newblock
\href{https://doi.org/10.1145/3359252}{doi:\nolinkurl{10.1145/3359252}}


\bibitem[Jones(1997)]%
        {jones_virtualcommunitiesvirtual_1997}
\bibfield{author}{\bibinfo{person}{Quentin Jones}.}
  \bibinfo{year}{1997}\natexlab{}.
\newblock \showarticletitle{Virtual-{{Communities}}, {{Virtual Settlements}} \&
  {{Cyber-Archaeology}}: A {{Theoretical Outline}}}.
\newblock \bibinfo{journal}{\emph{Journal of Computer-Mediated Communication}}
  \bibinfo{volume}{3}, \bibinfo{number}{3} (\bibinfo{date}{Dec.}
  \bibinfo{year}{1997}), \bibinfo{pages}{0--0}.
\newblock
\href{https://doi.org/10.1111/j.1083-6101.1997.tb00075.x}{doi:\nolinkurl{10.1111/j.1083-6101.1997.tb00075.x}}


\bibitem[Keegan and Fiesler(2017)]%
        {keegan_evolutionconsequences_2017}
\bibfield{author}{\bibinfo{person}{Brian Keegan} {and} \bibinfo{person}{Casey
  Fiesler}.} \bibinfo{year}{2017}\natexlab{}.
\newblock \showarticletitle{The Evolution and Consequences of Peer Producing
  Wikipedia's Rules}. In \bibinfo{booktitle}{\emph{Proceedings of the
  {{Eleventh International AAAI Conference}} on {{Web}} and {{Social Media}}}}.
  \bibinfo{publisher}{AAAI}, \bibinfo{address}{Montreal, Quebec, Canada},
  \bibinfo{pages}{112--121}.
\newblock


\bibitem[Kiene et~al\mbox{.}(2019)]%
        {kiene_technologicalframes_2019}
\bibfield{author}{\bibinfo{person}{Charles Kiene},
  \bibinfo{person}{Jialun~"Aaron" Jiang}, {and} \bibinfo{person}{Benjamin~Mako
  Hill}.} \bibinfo{year}{2019}\natexlab{}.
\newblock \showarticletitle{Technological Frames and User Innovation: Exploring
  Technological Change in Community Moderation Teams}.
\newblock \bibinfo{journal}{\emph{Proceedings of the ACM on Human-Computer
  Interaction: Computer Supported Cooperative Work}} \bibinfo{volume}{3},
  \bibinfo{number}{CSCW} (\bibinfo{date}{Nov.} \bibinfo{year}{2019}),
  \bibinfo{pages}{44:1--44:23}.
\newblock
\href{https://doi.org/10.1145/3359146}{doi:\nolinkurl{10.1145/3359146}}


\bibitem[Kiene et~al\mbox{.}(2016)]%
        {kiene_survivingeternal_2016}
\bibfield{author}{\bibinfo{person}{Charles Kiene}, \bibinfo{person}{Andr{\'e}s
  {Monroy-Hern{\'a}ndez}}, {and} \bibinfo{person}{Benjamin~Mako Hill}.}
  \bibinfo{year}{2016}\natexlab{}.
\newblock \showarticletitle{Surviving an ``{{Eternal September}}'': {{How}} an
  Online Community Managed a Surge of Newcomers}. In
  \bibinfo{booktitle}{\emph{Proceedings of the 2016 {{CHI Conference}} on
  {{Human Factors}} in {{Computing Systems}}}}. \bibinfo{publisher}{ACM},
  \bibinfo{address}{New York, NY}, \bibinfo{pages}{1152--1156}.
\newblock
\showISBNx{978-1-4503-3362-7}
\href{https://doi.org/10.1145/2858036.2858356}{doi:\nolinkurl{10.1145/2858036.2858356}}


\bibitem[Kraut et~al\mbox{.}(2012)]%
        {kraut_buildingsuccessful_2012}
\bibfield{author}{\bibinfo{person}{Robert~E. Kraut}, \bibinfo{person}{Paul
  Resnick}, \bibinfo{person}{Sara Kiesler}, \bibinfo{person}{Moira Burke},
  \bibinfo{person}{Yan Chen}, \bibinfo{person}{Niki Kittur},
  \bibinfo{person}{Joseph Konstan}, \bibinfo{person}{Yuqing Ren}, {and}
  \bibinfo{person}{John Riedl}.} \bibinfo{year}{2012}\natexlab{}.
\newblock \bibinfo{booktitle}{\emph{Building {{Successful Online Communities}}:
  {{Evidence-Based Social Design}}}}.
\newblock \bibinfo{publisher}{MIT Press}, \bibinfo{address}{Cambridge, MA}.
\newblock
\showISBNx{978-0-262-29831-5}
\href{https://doi.org/10.7551/mitpress/8472.001.0001}{doi:\nolinkurl{10.7551/mitpress/8472.001.0001}}


\bibitem[Kriplean et~al\mbox{.}(2007)]%
        {kriplean_communityconsensus_2007}
\bibfield{author}{\bibinfo{person}{Travis Kriplean}, \bibinfo{person}{Ivan
  Beschastnikh}, \bibinfo{person}{David~W. McDonald}, {and}
  \bibinfo{person}{Scott~A. Golder}.} \bibinfo{year}{2007}\natexlab{}.
\newblock \showarticletitle{Community, Consensus, Coercion, Control: Cs*w or
  How Policy Mediates Mass Participation}. In \bibinfo{booktitle}{\emph{Proc.
  of the 2007 {{International ACM Conference}} on {{Supporting Group Work}}}}.
  \bibinfo{publisher}{ACM}, \bibinfo{address}{New York, NY, USA},
  \bibinfo{pages}{167--176}.
\newblock
\showISBNx{978-1-59593-845-9}
\href{https://doi.org/10.1145/1316624.1316648}{doi:\nolinkurl{10.1145/1316624.1316648}}


\bibitem[La~Cava et~al\mbox{.}(2024)]%
        {lacava_polarizationdecentralized_2024}
\bibfield{author}{\bibinfo{person}{Lucio La~Cava}, \bibinfo{person}{Domenico
  Mandaglio}, {and} \bibinfo{person}{Andrea Tagarelli}.}
  \bibinfo{year}{2024}\natexlab{}.
\newblock \showarticletitle{Polarization in {{Decentralized Online Social
  Networks}}}. In \bibinfo{booktitle}{\emph{Proceedings of the 16th {{ACM Web
  Science Conference}}}} \emph{(\bibinfo{series}{{{WEBSCI}} '24})}.
  \bibinfo{publisher}{Association for Computing Machinery},
  \bibinfo{address}{New York, NY, USA}, \bibinfo{pages}{48--52}.
\newblock
\showISBNx{979-8-4007-0334-8}
\href{https://doi.org/10.1145/3614419.3644013}{doi:\nolinkurl{10.1145/3614419.3644013}}


\bibitem[Leimeister et~al\mbox{.}(2004)]%
        {leimeister_successfactors_2004}
\bibfield{author}{\bibinfo{person}{Jan~Marco Leimeister},
  \bibinfo{person}{Pascal Sidiras}, {and} \bibinfo{person}{Helmut Krcmar}.}
  \bibinfo{year}{2004}\natexlab{}.
\newblock \showarticletitle{Success Factors of Virtual Communities from the
  Perspective of Members and Operators: An Empirical Study}. In
  \bibinfo{booktitle}{\emph{Proceedings of the 37th {{Annual Hawaii
  International Conference}} on {{System Sciences}} ({{HICCS}})}}.
  \bibinfo{publisher}{IEEE}, \bibinfo{address}{Waikoloa Village, Hawaii, USA},
  \bibinfo{pages}{10}.
\newblock
\href{https://doi.org/10.1109/HICSS.2004.1265459}{doi:\nolinkurl{10.1109/HICSS.2004.1265459}}


\bibitem[Lin et~al\mbox{.}(2017)]%
        {lin_betterwhen_2017}
\bibfield{author}{\bibinfo{person}{Zhiyuan Lin}, \bibinfo{person}{Niloufar
  Salehi}, \bibinfo{person}{Bowen Yao}, \bibinfo{person}{Yiqi Chen}, {and}
  \bibinfo{person}{Michael~S. Bernstein}.} \bibinfo{year}{2017}\natexlab{}.
\newblock \showarticletitle{Better When It Was Smaller? {{Community}} Content
  and Behavior after Massive Growth.}. In \bibinfo{booktitle}{\emph{Eleventh
  {{International AAAI Conference}} on {{Web}} and {{Social Media}}}}.
  \bibinfo{publisher}{AAAI}, \bibinfo{address}{Montreal, Canada},
  \bibinfo{pages}{132--141}.
\newblock


\bibitem[March et~al\mbox{.}(2000)]%
        {march_dynamicsrules_2000}
\bibfield{author}{\bibinfo{person}{James March}, \bibinfo{person}{Martin
  Schulz}, {and} \bibinfo{person}{Xueguang Zhou}.}
  \bibinfo{year}{2000}\natexlab{}.
\newblock \bibinfo{booktitle}{\emph{The Dynamics of Rules: Change in Written
  Organizational Codes}}.
\newblock \bibinfo{publisher}{Stanford University Press},
  \bibinfo{address}{Stanford, CA}.
\newblock


\bibitem[Matei and Dobrescu(2010)]%
        {matei_wikipediasneutral_2010}
\bibfield{author}{\bibinfo{person}{Sorin~Adam Matei} {and}
  \bibinfo{person}{Caius Dobrescu}.} \bibinfo{year}{2010}\natexlab{}.
\newblock \showarticletitle{Wikipedia's ``Neutral Point of View'': Settling
  Conflict through Ambiguity}.
\newblock \bibinfo{journal}{\emph{The Information Society}}
  \bibinfo{volume}{27}, \bibinfo{number}{1} (\bibinfo{date}{Dec.}
  \bibinfo{year}{2010}), \bibinfo{pages}{40--51}.
\newblock
\showISSN{0197-2243}
\href{https://doi.org/10.1080/01972243.2011.534368}{doi:\nolinkurl{10.1080/01972243.2011.534368}}


\bibitem[Matias(2019a)]%
        {matias_civiclabor_2019}
\bibfield{author}{\bibinfo{person}{J.~Nathan Matias}.}
  \bibinfo{year}{2019}\natexlab{a}.
\newblock \showarticletitle{The Civic Labor of Volunteer Moderators Online}.
\newblock \bibinfo{journal}{\emph{Social Media + Society}} \bibinfo{volume}{5},
  \bibinfo{number}{2} (\bibinfo{date}{April} \bibinfo{year}{2019}),
  \bibinfo{pages}{1--12}.
\newblock
\showISSN{2056-3051, 2056-3051}
\href{https://doi.org/10.1177/2056305119836778}{doi:\nolinkurl{10.1177/2056305119836778}}


\bibitem[Matias(2019b)]%
        {matias_preventingharassment_2019}
\bibfield{author}{\bibinfo{person}{J.~Nathan Matias}.}
  \bibinfo{year}{2019}\natexlab{b}.
\newblock \showarticletitle{Preventing Harassment and Increasing Group
  Participation through Social Norms in 2,190 Online Science Discussions}.
\newblock \bibinfo{journal}{\emph{Proceedings of the National Academy of
  Sciences}} \bibinfo{volume}{116}, \bibinfo{number}{20} (\bibinfo{date}{May}
  \bibinfo{year}{2019}), \bibinfo{pages}{9785--9789}.
\newblock
\showISSN{0027-8424, 1091-6490}
\href{https://doi.org/10.1073/pnas.1813486116}{doi:\nolinkurl{10.1073/pnas.1813486116}}


\bibitem[McDonald et~al\mbox{.}(2019)]%
        {mcdonald_reliabilityinterrater_2019}
\bibfield{author}{\bibinfo{person}{Nora McDonald}, \bibinfo{person}{Sarita
  Schoenebeck}, {and} \bibinfo{person}{Andrea Forte}.}
  \bibinfo{year}{2019}\natexlab{}.
\newblock \showarticletitle{Reliability and {{Inter-rater Reliability}} in
  {{Qualitative Research}}: {{Norms}} and {{Guidelines}} for {{CSCW}} and {{HCI
  Practice}}}.
\newblock   \bibinfo{volume}{3} (\bibinfo{year}{2019}),
  \bibinfo{pages}{72:1--72:23}.
\newblock
Issue CSCW.
\href{https://doi.org/10.1145/3359174}{doi:\nolinkurl{10.1145/3359174}}


\bibitem[Meyer and Rowan(1977)]%
        {meyer_institutionalizedorganizations_1977}
\bibfield{author}{\bibinfo{person}{John~W. Meyer} {and} \bibinfo{person}{Brian
  Rowan}.} \bibinfo{year}{1977}\natexlab{}.
\newblock \showarticletitle{Institutionalized Organizations: {{Formal}}
  Structure as Myth and Ceremony}.
\newblock \bibinfo{journal}{\emph{Amer. J. Sociology}} \bibinfo{volume}{83},
  \bibinfo{number}{2} (\bibinfo{date}{Sept.} \bibinfo{year}{1977}),
  \bibinfo{pages}{340--363}.
\newblock
\showISSN{00029602}
\showeprint[jstor]{2778293}


\bibitem[Nicholson et~al\mbox{.}(2023)]%
        {nicholson_mastodonrules_2023}
\bibfield{author}{\bibinfo{person}{Matthew~N. Nicholson},
  \bibinfo{person}{Brian~C Keegan}, {and} \bibinfo{person}{Casey Fiesler}.}
  \bibinfo{year}{2023}\natexlab{}.
\newblock \showarticletitle{Mastodon {{Rules}}: {{Characterizing Formal Rules}}
  on {{Popular Mastodon Instances}}}. In \bibinfo{booktitle}{\emph{Companion
  {{Publication}} of the 2023 {{Conference}} on {{Computer Supported
  Cooperative Work}} and {{Social Computing}}}}
  \emph{(\bibinfo{series}{{{CSCW}} '23 {{Companion}}})}.
  \bibinfo{publisher}{Association for Computing Machinery},
  \bibinfo{address}{New York, NY, USA}, \bibinfo{pages}{86--90}.
\newblock
\showISBNx{979-8-4007-0129-0}
\href{https://doi.org/10.1145/3584931.3606970}{doi:\nolinkurl{10.1145/3584931.3606970}}


\bibitem[North(1990)]%
        {north_institutionsinstitutional_1990}
\bibfield{author}{\bibinfo{person}{Douglass~C North}.}
  \bibinfo{year}{1990}\natexlab{}.
\newblock \bibinfo{booktitle}{\emph{Institutions, Institutional Change, and
  Economic Performance}}.
\newblock \bibinfo{publisher}{Cambridge University Press},
  \bibinfo{address}{Cambridge, UK}.
\newblock
\showISBNx{978-0-511-80867-8}
\href{https://doi.org/10.1017/CBO9780511808678}{doi:\nolinkurl{10.1017/CBO9780511808678}}


\bibitem[Ostrom(1990)]%
        {ostrom_governingcommons_1990}
\bibfield{author}{\bibinfo{person}{Elinor Ostrom}.}
  \bibinfo{year}{1990}\natexlab{}.
\newblock \bibinfo{booktitle}{\emph{Governing the Commons: {{The}} Evolution of
  Institutions for Collective Action}}.
\newblock \bibinfo{publisher}{Cambridge University Press},
  \bibinfo{address}{Cambridge}.
\newblock
\href{https://doi.org/10.1017/CBO9780511807763}{doi:\nolinkurl{10.1017/CBO9780511807763}}


\bibitem[Ostrom(2005)]%
        {ostrom_understandinginstitutional_2005}
\bibfield{author}{\bibinfo{person}{Elinor Ostrom}.}
  \bibinfo{year}{2005}\natexlab{}.
\newblock \bibinfo{booktitle}{\emph{Understanding Institutional Diversity}}.
\newblock \bibinfo{publisher}{Princeton University Press},
  \bibinfo{address}{Princeton}.
\newblock
\showISBNx{978-0-691-12207-6}


\bibitem[Powell and Dimaggio(1991)]%
        {powell_newinstitutionalism_1991}
\bibfield{author}{\bibinfo{person}{Walter~W. Powell} {and}
  \bibinfo{person}{Paul Dimaggio}.} \bibinfo{year}{1991}\natexlab{}.
\newblock \bibinfo{booktitle}{\emph{The {{New Institutionalism}} in
  {{Organizational Analysis}}}}.
\newblock \bibinfo{publisher}{University of Chicago press},
  \bibinfo{address}{Chicago London}.
\newblock
\showISBNx{978-0-226-67708-8 978-0-226-67709-5}
\showLCCN{658.001}


\bibitem[Preece(2000)]%
        {preece_onlinecommunities_2000}
\bibfield{author}{\bibinfo{person}{Jenny Preece}.}
  \bibinfo{year}{2000}\natexlab{}.
\newblock \bibinfo{booktitle}{\emph{Online Communities: Designing Usability,
  Supporting Sociability}}.
\newblock \bibinfo{publisher}{John Wiley}, \bibinfo{address}{New York}.
\newblock
\showISBNx{978-0-471-80599-1}


\bibitem[Raman et~al\mbox{.}(2019)]%
        {raman_challengesdecentralised_2019}
\bibfield{author}{\bibinfo{person}{Aravindh Raman}, \bibinfo{person}{Sagar
  Joglekar}, \bibinfo{person}{Emiliano~De Cristofaro},
  \bibinfo{person}{Nishanth Sastry}, {and} \bibinfo{person}{Gareth Tyson}.}
  \bibinfo{year}{2019}\natexlab{}.
\newblock \showarticletitle{Challenges in the Decentralised Web: The Mastodon
  Case}. In \bibinfo{booktitle}{\emph{Proceedings of the {{Internet Measurement
  Conference}}}} \emph{(\bibinfo{series}{{{IMC}} '19})}.
  \bibinfo{publisher}{Association for Computing Machinery},
  \bibinfo{address}{New York, NY, USA}, \bibinfo{pages}{217--229}.
\newblock
\showISBNx{978-1-4503-6948-0}
\href{https://doi.org/10.1145/3355369.3355572}{doi:\nolinkurl{10.1145/3355369.3355572}}


\bibitem[Reddy and Chandrasekharan(2023)]%
        {reddy_evolutionrules_2023}
\bibfield{author}{\bibinfo{person}{Harita Reddy} {and} \bibinfo{person}{Eshwar
  Chandrasekharan}.} \bibinfo{year}{2023}\natexlab{}.
\newblock \showarticletitle{Evolution of {{Rules}} in {{Reddit Communities}}}.
  In \bibinfo{booktitle}{\emph{Companion {{Publication}} of the 2023
  {{Conference}} on {{Computer Supported Cooperative Work}} and {{Social
  Computing}}}} \emph{(\bibinfo{series}{{{CSCW}} '23 {{Companion}}})}.
  \bibinfo{publisher}{Association for Computing Machinery},
  \bibinfo{address}{New York, NY, USA}, \bibinfo{pages}{278--282}.
\newblock
\showISBNx{979-8-4007-0129-0}
\href{https://doi.org/10.1145/3584931.3606973}{doi:\nolinkurl{10.1145/3584931.3606973}}


\bibitem[Schneider(2022)]%
        {schneider_adminsmods_2022}
\bibfield{author}{\bibinfo{person}{Nathan Schneider}.}
  \bibinfo{year}{2022}\natexlab{}.
\newblock \showarticletitle{Admins, Mods, and Benevolent Dictators for Life:
  {{The}} Implicit Feudalism of Online Communities}.
\newblock \bibinfo{journal}{\emph{New Media \& Society}} \bibinfo{volume}{24},
  \bibinfo{number}{9} (\bibinfo{date}{Sept.} \bibinfo{year}{2022}),
  \bibinfo{pages}{1965--1985}.
\newblock
\showISSN{1461-4448}
\href{https://doi.org/10.1177/1461444820986553}{doi:\nolinkurl{10.1177/1461444820986553}}


\bibitem[{Sch{\"o}pke-Gonzalez} et~al\mbox{.}(2024)]%
        {schopke-gonzalez_whyvolunteer_2024}
\bibfield{author}{\bibinfo{person}{Angela~M. {Sch{\"o}pke-Gonzalez}},
  \bibinfo{person}{Shubham Atreja}, \bibinfo{person}{Han~Na Shin},
  \bibinfo{person}{Najmin Ahmed}, {and} \bibinfo{person}{Libby Hemphill}.}
  \bibinfo{year}{2024}\natexlab{}.
\newblock \showarticletitle{Why Do Volunteer Content Moderators Quit?
  {{Burnout}}, Conflict, and Harmful Behaviors}.
\newblock \bibinfo{journal}{\emph{New Media \& Society}} \bibinfo{volume}{26},
  \bibinfo{number}{10} (\bibinfo{date}{Oct.} \bibinfo{year}{2024}),
  \bibinfo{pages}{5677--5701}.
\newblock
\showISSN{1461-4448}
\href{https://doi.org/10.1177/14614448221138529}{doi:\nolinkurl{10.1177/14614448221138529}}


\bibitem[Seering(2019)]%
        {seering_buildingmore_2019}
\bibfield{author}{\bibinfo{person}{Joseph Seering}.}
  \bibinfo{year}{2019}\natexlab{}.
\newblock \showarticletitle{Building {{More Positive Online Communities}}
  through {{Improving Moderation}} and {{Strengthening Social Identity}}}. In
  \bibinfo{booktitle}{\emph{Conference {{Companion Publication}} of the 2019 on
  {{Computer Supported Cooperative Work}} and {{Social Computing}}}}
  \emph{(\bibinfo{series}{{{CSCW}} '19})}. \bibinfo{publisher}{Association for
  Computing Machinery}, \bibinfo{address}{New York, NY, USA},
  \bibinfo{pages}{89--93}.
\newblock
\showISBNx{978-1-4503-6692-2}
\href{https://doi.org/10.1145/3311957.3361855}{doi:\nolinkurl{10.1145/3311957.3361855}}


\bibitem[Seering and Kairam(2022)]%
        {seering_whomoderates_2022}
\bibfield{author}{\bibinfo{person}{Joseph Seering} {and}
  \bibinfo{person}{Sanjay~R. Kairam}.} \bibinfo{year}{2022}\natexlab{}.
\newblock \showarticletitle{Who {{Moderates}} on {{Twitch}} and {{What Do They
  Do}}? {{Quantifying Practices}} in {{Community Moderation}} on {{Twitch}}}.
\newblock \bibinfo{journal}{\emph{Proc. ACM Hum.-Comput. Interact.}}
  \bibinfo{volume}{7}, \bibinfo{number}{GROUP} (\bibinfo{date}{Dec.}
  \bibinfo{year}{2022}), \bibinfo{pages}{18:1--18:18}.
\newblock
\href{https://doi.org/10.1145/3567568}{doi:\nolinkurl{10.1145/3567568}}


\bibitem[Seering et~al\mbox{.}(2020)]%
        {seering_metaphorsmoderation_2020}
\bibfield{author}{\bibinfo{person}{Joseph Seering}, \bibinfo{person}{Geoff
  Kaufman}, {and} \bibinfo{person}{Stevie Chancellor}.}
  \bibinfo{year}{2020}\natexlab{}.
\newblock \showarticletitle{Metaphors in Moderation}.
\newblock \bibinfo{journal}{\emph{New Media \& Society}} \bibinfo{volume}{24},
  \bibinfo{number}{3} (\bibinfo{date}{Oct.} \bibinfo{year}{2020}),
  \bibinfo{pages}{1--20}.
\newblock
\showISSN{1461-4448}
\href{https://doi.org/10.1177/1461444820964968}{doi:\nolinkurl{10.1177/1461444820964968}}


\bibitem[Seering et~al\mbox{.}(2017)]%
        {seering_shapingpro_2017}
\bibfield{author}{\bibinfo{person}{Joseph Seering}, \bibinfo{person}{Robert
  Kraut}, {and} \bibinfo{person}{Laura Dabbish}.}
  \bibinfo{year}{2017}\natexlab{}.
\newblock \showarticletitle{Shaping {{Pro}} and {{Anti-Social Behavior}} on
  {{Twitch Through Moderation}} and {{Example-Setting}}}. In
  \bibinfo{booktitle}{\emph{Proceedings of the 2017 {{ACM Conference}} on
  {{Computer Supported Cooperative Work}} and {{Social Computing}}}}
  \emph{(\bibinfo{series}{{{CSCW}} '17})}. \bibinfo{publisher}{ACM},
  \bibinfo{address}{New York, NY, USA}, \bibinfo{pages}{111--125}.
\newblock
\showISBNx{978-1-4503-4335-0}
\href{https://doi.org/10.1145/2998181.2998277}{doi:\nolinkurl{10.1145/2998181.2998277}}


\bibitem[Seering et~al\mbox{.}(2019)]%
        {seering_moderatorengagement_2019}
\bibfield{author}{\bibinfo{person}{Joseph Seering}, \bibinfo{person}{Tony
  Wang}, \bibinfo{person}{Jina Yoon}, {and} \bibinfo{person}{Geoff Kaufman}.}
  \bibinfo{year}{2019}\natexlab{}.
\newblock \showarticletitle{Moderator Engagement and Community Development in
  the Age of Algorithms}.
\newblock \bibinfo{journal}{\emph{New Media \& Society}} \bibinfo{volume}{21},
  \bibinfo{number}{7} (\bibinfo{date}{Jan.} \bibinfo{year}{2019}),
  \bibinfo{pages}{1--24}.
\newblock
\showISSN{1461-4448}
\href{https://doi.org/10.1177/1461444818821316}{doi:\nolinkurl{10.1177/1461444818821316}}


\bibitem[Shaw and Hill(2014)]%
        {shaw_laboratoriesoligarchy_2014}
\bibfield{author}{\bibinfo{person}{Aaron Shaw} {and}
  \bibinfo{person}{Benjamin~Mako Hill}.} \bibinfo{year}{2014}\natexlab{}.
\newblock \showarticletitle{Laboratories of Oligarchy? {{How}} the Iron Law
  Extends to Peer Production}.
\newblock \bibinfo{journal}{\emph{Journal of Communication}}
  \bibinfo{volume}{64}, \bibinfo{number}{2} (\bibinfo{year}{2014}),
  \bibinfo{pages}{215--238}.
\newblock
\showISSN{1460-2466}
\href{https://doi.org/10.1111/jcom.12082}{doi:\nolinkurl{10.1111/jcom.12082}}


\bibitem[Singh et~al\mbox{.}(1986)]%
        {singh_organizationallegitimacy_1986}
\bibfield{author}{\bibinfo{person}{Jitendra~V. Singh},
  \bibinfo{person}{David~J. Tucker}, {and} \bibinfo{person}{Robert~J. House}.}
  \bibinfo{year}{1986}\natexlab{}.
\newblock \showarticletitle{Organizational Legitimacy and the Liability of
  Newness}.
\newblock \bibinfo{journal}{\emph{Administrative Science Quarterly}}
  \bibinfo{volume}{31}, \bibinfo{number}{2} (\bibinfo{year}{1986}),
  \bibinfo{pages}{171--193}.
\newblock
\showISSN{0001-8392}
\href{https://doi.org/10.2307/2392787}{doi:\nolinkurl{10.2307/2392787}}
\showeprint[jstor]{2392787}


\bibitem[Stockinger et~al\mbox{.}(2023)]%
        {stockinger_navigatinggray_2023}
\bibfield{author}{\bibinfo{person}{Andrea Stockinger}, \bibinfo{person}{Svenja
  Sch{\"a}fer}, {and} \bibinfo{person}{Sophie Lecheler}.}
  \bibinfo{year}{2023}\natexlab{}.
\newblock \showarticletitle{Navigating the Gray Areas of Content Moderation:
  {{Professional}} Moderators' Perspectives on Uncivil User Comments and the
  Role of ({{AI-based}}) Technological Tools}.
\newblock \bibinfo{journal}{\emph{New Media \& Society}} \bibinfo{volume}{27},
  \bibinfo{number}{3} (\bibinfo{date}{Aug.} \bibinfo{year}{2023}),
  \bibinfo{pages}{14614448231190901}.
\newblock
\showISSN{1461-4448}
\href{https://doi.org/10.1177/14614448231190901}{doi:\nolinkurl{10.1177/14614448231190901}}


\bibitem[Tan(2018)]%
        {tan_tracingcommunity_2018}
\bibfield{author}{\bibinfo{person}{Chenhao Tan}.}
  \bibinfo{year}{2018}\natexlab{}.
\newblock \showarticletitle{Tracing Community Genealogy: How New Communities
  Emerge from the Old}. In \bibinfo{booktitle}{\emph{Proceedings of the
  {{Twelfth International Conference}} on {{Web}} and {{Social Media}}
  ({{ICWSM}} '18)}}. \bibinfo{publisher}{AAAI}, \bibinfo{address}{Palo Alto,
  California}, \bibinfo{pages}{395--404}.
\newblock


\bibitem[TeBlunthuis(2025)]%
        {teblunthuis_nichedynamics_2025}
\bibfield{author}{\bibinfo{person}{Nathan TeBlunthuis}.}
  \bibinfo{year}{2025}\natexlab{}.
\newblock \showarticletitle{Niche {{Dynamics}} in {{Complex Online Community
  Ecosystems}}}.
\newblock \bibinfo{journal}{\emph{Proceedings of the International AAAI
  Conference on Web and Social Media}}  \bibinfo{volume}{19}
  (\bibinfo{date}{June} \bibinfo{year}{2025}), \bibinfo{pages}{1880--1892}.
\newblock
\showISSN{2334-0770}
\href{https://doi.org/10.1609/icwsm.v19i1.35907}{doi:\nolinkurl{10.1609/icwsm.v19i1.35907}}
\showeprint[arxiv]{2504.02153}~[cs]


\bibitem[TeBlunthuis and Hill(2022)]%
        {teblunthuis_identifyingcompetition_2022}
\bibfield{author}{\bibinfo{person}{Nathan TeBlunthuis} {and}
  \bibinfo{person}{Benjamin~Mako Hill}.} \bibinfo{year}{2022}\natexlab{}.
\newblock \showarticletitle{Identifying Competition and Mutualism between
  Online Groups}. In \bibinfo{booktitle}{\emph{International {{AAAI
  Conference}} on {{Web}} and {{Social Media}} ({{ICWSM}} 2022)}},
  Vol.~\bibinfo{volume}{16}. \bibinfo{publisher}{AAAI},
  \bibinfo{address}{Atlanta, Georgia, USA}, \bibinfo{pages}{993--1004}.
\newblock


\bibitem[TeBlunthuis et~al\mbox{.}(2021)]%
        {teblunthuis_effectsalgorithmic_2021}
\bibfield{author}{\bibinfo{person}{Nathan TeBlunthuis},
  \bibinfo{person}{Benjamin~Mako Hill}, {and} \bibinfo{person}{Aaron
  Halfaker}.} \bibinfo{year}{2021}\natexlab{}.
\newblock \showarticletitle{Effects of {{Algorithmic Flagging}} on
  {{Fairness}}: {{Quasi-experimental Evidence}} from {{Wikipedia}}}.
\newblock \bibinfo{journal}{\emph{Proceedings of the ACM on Human-Computer
  Interaction}} \bibinfo{volume}{5}, \bibinfo{number}{CSCW1}
  (\bibinfo{date}{April} \bibinfo{year}{2021}), \bibinfo{pages}{56:1--56:27}.
\newblock
\href{https://doi.org/10.1145/3449130}{doi:\nolinkurl{10.1145/3449130}}


\bibitem[TeBlunthuis et~al\mbox{.}(2022)]%
        {teblunthuis_nocommunity_2022}
\bibfield{author}{\bibinfo{person}{Nathan TeBlunthuis},
  \bibinfo{person}{Charles Kiene}, \bibinfo{person}{Isabella Brown},
  \bibinfo{person}{Laura~(Alia) Levi}, \bibinfo{person}{Nicole McGinnis}, {and}
  \bibinfo{person}{Benjamin~Mako Hill}.} \bibinfo{year}{2022}\natexlab{}.
\newblock \showarticletitle{No Community Can Do Everything: Why People
  Participate in Similar Online Communities}.
\newblock \bibinfo{journal}{\emph{Proceedings of the ACM on Human-Computer
  Interaction: Computer Supported Cooperative Work}}  \bibinfo{volume}{6}
  (\bibinfo{date}{April} \bibinfo{year}{2022}), \bibinfo{pages}{1--25}.
\newblock
\href{https://doi.org/10.1145/3512908}{doi:\nolinkurl{10.1145/3512908}}


\bibitem[TeBlunthuis et~al\mbox{.}(2018)]%
        {teblunthuis_revisitingrise_2018}
\bibfield{author}{\bibinfo{person}{Nathan TeBlunthuis}, \bibinfo{person}{Aaron
  Shaw}, {and} \bibinfo{person}{Benjamin~Mako Hill}.}
  \bibinfo{year}{2018}\natexlab{}.
\newblock \showarticletitle{Revisiting "{{The}} Rise and Decline" in a
  Population of Peer Production Projects}. In
  \bibinfo{booktitle}{\emph{Proceedings of the 2018 {{CHI Conference}} on
  {{Human Factors}} in {{Computing Systems}}}}. \bibinfo{publisher}{ACM},
  \bibinfo{address}{New York, NY}, \bibinfo{pages}{355:1--355:7}.
\newblock
\showISBNx{978-1-4503-5620-6}
\href{https://doi.org/10.1145/3173574.3173929}{doi:\nolinkurl{10.1145/3173574.3173929}}


\bibitem[Tolbert and Zucker(1983)]%
        {tolbert_institutionalsources_1983}
\bibfield{author}{\bibinfo{person}{Pamela~S. Tolbert} {and}
  \bibinfo{person}{Lynne~G. Zucker}.} \bibinfo{year}{1983}\natexlab{}.
\newblock \showarticletitle{Institutional Sources of Change in the Formal
  Structure of Organizations: The Diffusion of Civil Service Reform,
  1880-1935}.
\newblock \bibinfo{journal}{\emph{Administrative Science Quarterly}}
  \bibinfo{volume}{28}, \bibinfo{number}{1} (\bibinfo{year}{1983}),
  \bibinfo{pages}{22--39}.
\newblock
\showISSN{0001-8392}
\href{https://doi.org/10.2307/2392383}{doi:\nolinkurl{10.2307/2392383}}
\showeprint[jstor]{2392383}


\bibitem[Tosch et~al\mbox{.}(2024)]%
        {tosch_privacypolicies_2024}
\bibfield{author}{\bibinfo{person}{Emma Tosch}, \bibinfo{person}{Luis Garcia},
  \bibinfo{person}{Cynthia Li}, {and} \bibinfo{person}{Chris Martens}.}
  \bibinfo{year}{2024}\natexlab{}.
\newblock \showarticletitle{Privacy {{Policies}} on the {{Fediverse}}: {{A Case
  Study}} of {{Mastodon Instances}}}.
\newblock \bibinfo{journal}{\emph{Proceedings on Privacy Enhancing
  Technologies}} \bibinfo{volume}{2024}, \bibinfo{number}{4}
  (\bibinfo{date}{Oct.} \bibinfo{year}{2024}), \bibinfo{pages}{700--733}.
\newblock
\showISSN{2299-0984}
\href{https://doi.org/10.56553/popets-2024-0138}{doi:\nolinkurl{10.56553/popets-2024-0138}}


\bibitem[Turco(2016)]%
        {turco_conversationalfirm_2016}
\bibfield{author}{\bibinfo{person}{Catherine~J Turco}.}
  \bibinfo{year}{2016}\natexlab{}.
\newblock \bibinfo{booktitle}{\emph{The Conversational Firm: {{Rethinking}}
  Bureaucracy in the Age of Social Media}}.
\newblock \bibinfo{publisher}{Columbia University Press}, \bibinfo{address}{New
  York, NY}.
\newblock
\showISBNx{978-0-231-17898-3}


\bibitem[Weld et~al\mbox{.}(2021)]%
        {weld_whatmakes_2021}
\bibfield{author}{\bibinfo{person}{Galen Weld}, \bibinfo{person}{Amy~X. Zhang},
  {and} \bibinfo{person}{Tim Althoff}.} \bibinfo{year}{2021}\natexlab{}.
\newblock \showarticletitle{What Makes Online Communities 'Better'?
  {{Measuring}} Values, Consensus, and Conflict across Thousands of
  Subreddits}.
\newblock \bibinfo{journal}{\emph{arXiv:2111.05835 [cs]}}  \bibinfo{volume}{16}
  (\bibinfo{date}{Nov.} \bibinfo{year}{2021}), \bibinfo{pages}{1121--1132}.
\newblock
\showeprint[arxiv]{2111.05835}~[cs]


\bibitem[Wohn(2019)]%
        {wohn_volunteermoderators_2019}
\bibfield{author}{\bibinfo{person}{Yvette Wohn}.}
  \bibinfo{year}{2019}\natexlab{}.
\newblock \showarticletitle{Volunteer Moderators in Twitch Micro Communities:
  How They Get Involved, the Roles They Play, and the Emotional Labor They
  Experience}. In \bibinfo{booktitle}{\emph{Proceedings of the 2019 {{CHI
  Conference}} on {{Human Factors}} in {{Computing Systems}}}}
  \emph{(\bibinfo{series}{{{CHI}}'19})}. \bibinfo{publisher}{ACM},
  \bibinfo{address}{Glasgow, UK}, \bibinfo{pages}{1--13}.
\newblock


\bibitem[Wu et~al\mbox{.}(2025)]%
        {wu_aididnt_2025}
\bibfield{author}{\bibinfo{person}{Yiwei Wu}, \bibinfo{person}{Leah Ajmani},
  \bibinfo{person}{Nathan TeBlunthuis}, {and} \bibinfo{person}{Hanlin Li}.}
  \bibinfo{year}{2025}\natexlab{}.
\newblock \bibinfo{booktitle}{\emph{{{AI Didn}}'t {{Start}} the {{Fire}}:
  {{Examining}} the {{Stack Exchange Moderator}} and {{Contributor Strike}}}}.
\newblock
\href{https://doi.org/10.48550/arXiv.2512.08884}{doi:\nolinkurl{10.48550/arXiv.2512.08884}}
\showeprint[arXiv]{2512.08884}~[cs]


\bibitem[Young(1988)]%
        {young_populationecology_1988}
\bibfield{author}{\bibinfo{person}{Ruth~C. Young}.}
  \bibinfo{year}{1988}\natexlab{}.
\newblock \showarticletitle{Is Population Ecology a Useful Paradigm for the
  Study of Organizations?}
\newblock \bibinfo{journal}{\emph{Amer. J. Sociology}} \bibinfo{volume}{94},
  \bibinfo{number}{7} (\bibinfo{year}{1988}), \bibinfo{pages}{1--24}.
\newblock
\showISSN{0002-9602}
\showeprint[jstor]{2781020}


\bibitem[Zhang et~al\mbox{.}(2020)]%
        {zhang_policykitbuilding_2020}
\bibfield{author}{\bibinfo{person}{Amy~X. Zhang}, \bibinfo{person}{Grant Hugh},
  {and} \bibinfo{person}{Michael~S. Bernstein}.}
  \bibinfo{year}{2020}\natexlab{}.
\newblock \showarticletitle{{{PolicyKit}}: {{Building}} Governance in Online
  Communities}. In \bibinfo{booktitle}{\emph{Proceedings of the 33rd {{Annual
  ACM Symposium}} on {{User Interface Software}} and {{Technology}}}}.
  \bibinfo{publisher}{ACM}, \bibinfo{address}{Virtual Event USA},
  \bibinfo{pages}{365--378}.
\newblock
\href{https://doi.org/10.1145/3379337.3415858}{doi:\nolinkurl{10.1145/3379337.3415858}}
\showeprint[arxiv]{2008.04236}


\bibitem[Zhang et~al\mbox{.}(2024)]%
        {zhang_troubleparadise_2024}
\bibfield{author}{\bibinfo{person}{Zhilin Zhang}, \bibinfo{person}{Jun Zhao},
  \bibinfo{person}{Ge Wang}, \bibinfo{person}{Samantha-Kaye Johnston},
  \bibinfo{person}{George Chalhoub}, \bibinfo{person}{Tala Ross},
  \bibinfo{person}{Diyi Liu}, \bibinfo{person}{Claudine Tinsman},
  \bibinfo{person}{Rui Zhao}, \bibinfo{person}{Max Van~Kleek}, {and}
  \bibinfo{person}{Nigel Shadbolt}.} \bibinfo{year}{2024}\natexlab{}.
\newblock \showarticletitle{Trouble in {{Paradise}}? {{Understanding Mastodon
  Admin}}'s {{Motivations}}, {{Experiences}}, and {{Challenges Running
  Decentralised Social Media}}}.
\newblock \bibinfo{journal}{\emph{Proc. ACM Hum.-Comput. Interact.}}
  \bibinfo{volume}{8}, \bibinfo{number}{CSCW2} (\bibinfo{date}{Nov.}
  \bibinfo{year}{2024}), \bibinfo{pages}{520:1--520:24}.
\newblock
\href{https://doi.org/10.1145/3687059}{doi:\nolinkurl{10.1145/3687059}}


\bibitem[Zhu et~al\mbox{.}(2013)]%
        {zhu_effectivenessshared_2013}
\bibfield{author}{\bibinfo{person}{Haiyi Zhu}, \bibinfo{person}{Robert~E.
  Kraut}, {and} \bibinfo{person}{Aniket Kittur}.}
  \bibinfo{year}{2013}\natexlab{}.
\newblock \showarticletitle{Effectiveness of {{Shared Leadership}} in
  {{Wikipedia}}}.
\newblock \bibinfo{journal}{\emph{Human Factors: The Journal of the Human
  Factors and Ergonomics Society}} \bibinfo{volume}{55}, \bibinfo{number}{6}
  (\bibinfo{date}{Dec.} \bibinfo{year}{2013}), \bibinfo{pages}{1021--1043}.
\newblock
\showISSN{0018-7208, 1547-8181}
\href{https://doi.org/10.1177/0018720813515704}{doi:\nolinkurl{10.1177/0018720813515704}}


\bibitem[Zhu et~al\mbox{.}(2014)]%
        {zhu_impactmembership_2014}
\bibfield{author}{\bibinfo{person}{Haiyi Zhu}, \bibinfo{person}{Robert~E.
  Kraut}, {and} \bibinfo{person}{Aniket Kittur}.}
  \bibinfo{year}{2014}\natexlab{}.
\newblock \showarticletitle{The Impact of Membership Overlap on the Survival of
  Online Communities}. In \bibinfo{booktitle}{\emph{Proceedings of the {{SIGCHI
  Conference}} on {{Human Factors}} in {{Computing Systems}}}}
  \emph{(\bibinfo{series}{{{CHI}} '14})}. \bibinfo{publisher}{Association for
  Computing Machinery}, \bibinfo{address}{New York, NY, USA},
  \bibinfo{pages}{281--290}.
\newblock
\showISBNx{978-1-4503-2473-1}
\href{https://doi.org/10.1145/2556288.2557213}{doi:\nolinkurl{10.1145/2556288.2557213}}


\bibitem[Zulli et~al\mbox{.}(2020)]%
        {zulli_rethinkingsocial_2020}
\bibfield{author}{\bibinfo{person}{Diana Zulli}, \bibinfo{person}{Miao Liu},
  {and} \bibinfo{person}{Robert Gehl}.} \bibinfo{year}{2020}\natexlab{}.
\newblock \showarticletitle{Rethinking the ``Social'' in ``Social Media'':
  {{Insights}} into Topology, Abstraction, and Scale on the {{Mastodon}} Social
  Network}.
\newblock \bibinfo{journal}{\emph{New Media \& Society}} \bibinfo{volume}{22},
  \bibinfo{number}{7} (\bibinfo{date}{July} \bibinfo{year}{2020}),
  \bibinfo{pages}{1188--1205}.
\newblock
\showISSN{1461-4448}
\href{https://doi.org/10.1177/1461444820912533}{doi:\nolinkurl{10.1177/1461444820912533}}


\end{thebibliography}

\clearpage
\newpage

\appendix
\section*{Appendices}

\section{Community Sampling Methods}
\label{appendix:sampling}
To identify and select interview participants across Reddit, Fandom, and the Fediverse, we implemented a multi-step sampling strategy focused on communities with similar rules. This process involved 1) compiling datasets of rule sets for each of the three platforms, 2) conducting a clustering analysis based on rule text, and 3) manually analyzing communities within rule similarity clusters to select samples of community leaders for interview recruitment.

\subsection*{Data Collection}
Three members of the research team---a postdoctoral fellow and two graduate students---constructed datasets capturing community rules on each platform.
\begin{itemize}
    \item \textbf{Reddit:} We crawled Reddit's public API from 2020 to 2022 and recorded every publicly available subreddit's rules data and community characteristics. We excluded subreddits with fewer than 1,000 members to ensure some expectation of activity, resulting in a final dataset with approximately 700,000 communities.
    \item \textbf{Fandom:} We used a February 2020 snapshot from the Internet Archive\footnote{\url{https://archive.org/details/wikiteam_2020-02-09}} that included the edit histories of approximately 340,000 Fandom wikis. We excluded communities with fewer than 200 unique article contributors and 10,000 article edits to ensure activity as well. We then downloaded policy and rule-related pages for each community.
    \item \textbf{Fediverse:} We used an iterative sampling process beginning with the server \texttt{mastodon.social} (because of its centrality and status within the broader Fediverse network), discovering additional servers from its peer connections, and continuing to discover connections from connections. From this list of Fediverse servers, we then downloaded and extracted the rule texts from the \texttt{/about/more} HTML data on Mastodon servers and the \texttt{/static/terms-of-service.html} HTML data on Pleroma servers as a snapshot of data from September 25, 2021.
\end{itemize}

\subsection*{Cluster Analysis}
To identify clusters of communities with similar rule language:
\begin{itemize}
    \item \textbf{Reddit and Fandom:} We created a 2-dimensional encoding of each community's rules by concatenating their rule texts, reducing the dimensionality via the Doc2Vec algorithm using the \texttt{gensim} Python package, and then further reducing the dimensionality using the \texttt{UMAP} Python package. We then used the \texttt{HDBSCAN} algorithm to group similar communities into clusters, giving us clusters of communities with similar rule language.
    \item \textbf{Fediverse:} For Fediverse communities, we generated a list of clusters of communities with similar rule language using the \texttt{HDBSCAN} algorithm after we computed pairwise n-gram similarity of communities' rule texts using the following formula:
    \begin{center}
    $\frac{\text{count}(X_1 \cap X_2)}{\min\left(\text{count}(X_1), \text{count}(X_2)\right)}$    
    \end{center}
\end{itemize}

\subsection*{Sample Selection}
For two weeks in June 2022, we assigned each empirical setting (Reddit, Fandom, and the Fediverse) to a subteam comprising three undergraduate RAs and at least one graduate or post-doc researcher. To reduce the scope of recruiting efforts, these subteams evaluated the clusters to narrow down potential communities from which to sample leaders for interviews. Subteams analyzed clusters of varying sizes (e.g., small, medium, and large) based on:
\begin{itemize}
    \item Recent activity levels
    \item Estimated community size
    \item Thematic content
    \item Rule characteristics
\end{itemize}

Subteams met weekly to discuss the results and created lists of communities to target for recruiting leaders for interviews.

\clearpage
\newpage
\section{Interview Questionnaire}
\label{appendix:protocol}
\subsection*{Questions about participation}
\begin{enumerate}
    \item What brought you to community \textless Community A\textgreater{}?
    \item How did you choose to participate in \textless Community A\textgreater{}?
    \item How long have you been involved with \textless Community A\textgreater{}?
    \item How do you see your role within \textless Community A\textgreater{}?
    \item How would you describe \textless Community A\textgreater{}?
    \item Are you paid or otherwise compensated for your participation in \textless Community A\textgreater{}?
    \item Do \textless Community A\textgreater{}'s choice of rules influence your decisions to choose to join or participate in \textless Community A\textgreater{} in any way? How so?
    \item Have you ever had a negative experience with \textless Community A\textgreater{}? If so, was it related to the rules in any way?
\end{enumerate}

\subsection*{Questions about rules in general}

\begin{enumerate}
    \item Can you tell me about \textless Community A\textgreater{}'s rules?
    \item What do you believe is the purpose of rules in \textless Community A\textgreater{}?
    \item Can you confirm that \textless Community A\textgreater{} uses the rules on their website?
    \item Were you aware of \textless Community A\textgreater{}'s rules before you signed up?
    \item Can you recall some specific rules of \textless Community A\textgreater{}? Why do these ones stick out to you?
    \item How important are the rules in \textless Community A\textgreater{}?
    \item What do you think would happen if you didn't have \textless rule X\textgreater{}?
    \item Are there any unwritten rules or norms in \textless Community A\textgreater{}?
    \item Are there any rules that \textless Community A\textgreater{} has that you wish it didn't? If so, why do you think they still have this rule?
    \item How do newcomers learn the rules of \textless Community A\textgreater{}? How did you learn of the rules of \textless Community A\textgreater{} when you first joined?
    \item Can you walk me through the first time you interacted with \textless Community A\textgreater{}'s rules / how you became familiar with \textless Community A\textgreater{}'s rules?
    \item What kinds of behaviors does \textless Community A\textgreater{} view negatively? Why?
    \item What kinds of behaviors does \textless Community A\textgreater{} view positively? Why?
    \item (In follow up to the two questions above) How do \textless Community A\textgreater{}'s rules address \textless Behavior Y\textgreater{}?
    \item Are there any rules that you wish that \textless Community A\textgreater{} had? Why? If so, why do you think the community doesn't have a rule like that?
    \item How do \textless Community A\textgreater{}'s rules align with your personal values? How do these rules reflect your values or not?
    \item What are your personal values in community participation?
    \item Can you tell me about your personal approaches to community participation?
    \item How do \textless Community A\textgreater{}'s rules shape your contributions to this community? For example, do the rules affect the types of content or work that you produce and/or share?
    \item In what types of settings or activities are your rules applied? Do the same rules apply in the mailing list as in the chat room? Do the same rules apply if you meet up in-person or at a convention?
    \item Do you have different rules for different mediums or contexts? (e.g., mailing lists, chat rooms, forums, in-person meetups)
    \item Are there things that tend to be agreed upon within \textless Community A\textgreater{} that aren't written in formal rules?
    \item How do the rules help, hurt, or otherwise impact the community?
    \item Can you describe any problems in \textless Community A\textgreater{}?
    \item Do the rules attempt to deal with these problems in any way?
\end{enumerate}

\subsection*{Questions about the origins of rules}

\begin{enumerate}
    \item How did \textless Community A\textgreater{} choose its rules?
    \item Why did you choose these rules?
    \item Have you ever had to change the rules?
    \item How have the rules changed, if ever?
    \item Are there any other communities that \textless Community A\textgreater{} looked to when you were choosing rules?
    \item What is the most recent rule that \textless Community A\textgreater{} has adopted? Tell me about that?
    \item How do you think \textless Community A\textgreater{} decided to phrase \textless Rule X\textgreater{}?
    \item Are there any rules that \textless Community A\textgreater{} used to have, but doesn't have anymore? If so, do you know why this change was made?
    \item What is the process for adopting or changing rules in \textless Community A\textgreater{}?
    \item Do you know who set the rules in \textless Community A\textgreater{}? Can you tell me more about that?
    \item Where do [you think] these rules come from? (looking at/through the rule list together)
\end{enumerate}

\subsection*{Questions for moderators, administrators, or other leadership positions}

\begin{enumerate}
    \item Is moderating \textless Community A\textgreater{} part of your job or otherwise a paid activity?
    \item Why do you work as a moderator of \textless Community A\textgreater{}?
    \item How did you become a moderator of \textless Community A\textgreater{}?
    \item How would you describe your work as a moderator of \textless Community A\textgreater{}?
\end{enumerate}

\subsection*{Questions about the enforcement of rules}

\begin{enumerate}
    \item What do you think \textless rule X\textgreater{} means in practice? How does that feel different to what the rule means in theory, if at all?
    \item How many people are involved in rule enforcement? What roles do they hold within the community?
    \item Do you feel as though the rules are fairly/equally enforced in \textless Community A\textgreater{}? (Follow up: why do you think that?)
    \item Do you feel as though the rules are consistently enforced in \textless Community A\textgreater{}? (Follow up: why do you think that?)
    \item Do you feel as though it is possible for rules to be effectively enforced in \textless Community A\textgreater{}? (Follow up: why do you think that?)
    \item How do you think moderators/administrators of \textless Community A\textgreater{} use the rules?
    \item How strictly do the moderators of \textless Community A\textgreater{} enforce the rules? Why? What makes you feel like it is strict or not?
    \item Do ordinary members of the community help enforce the rules or norms? (e.g., ``call each other out'')? If so, can you share an example of a time that happened?
    \item If not, why do you think not?
    \item In what ways are rules or norms enforced in \textless Community A\textgreater{}? Can you share an example of a time that happened?
    \item Are there particular people who take responsibility for enforcing the rules or norms? Are these people normally in formal roles, like moderators or administrators?
    \item What happens when there is conflict over how rules are enforced (internal to rules enforcement team)?
    \item What happens when there is conflict over how rules are enforced (external to rules enforcement team/community wide)?
    \item Are there any rules for moderators specifically?
\end{enumerate}

\subsection*{Questions about related communities}

\begin{enumerate}
    \item Are you a part of other online communities?
    \item How do the rules of \textless Community A\textgreater{} compare to the rules of related communities?
    \item Do all communities like yours have a rule like \textless rule X\textgreater{}?
    \item Can you name some other communities with rules?
    \item Can you compare the rules of \textless Community A\textgreater{} to \textless Community B\textgreater{}?
    \item How do the problems of \textless Community A\textgreater{} compare to those of \textless Community B\textgreater{}?
    \item Do you think differences in rules shape the relationship between \textless Community A\textgreater{} and \textless Community B\textgreater{} in any way?
\end{enumerate}

\subsection*{Questions about rules and platforms}

\begin{enumerate}
    \item Do you feel like you need to have certain rules because you are on \textless Platform A\textgreater{}?
    \item How important are the rules on \textless Platform A\textgreater{}?
    \item Can you tell me about the [most important] rules on \textless Platform A\textgreater{}? How does that translate into \textless Community A\textgreater{}?
    \item How do \textless Community A\textgreater{}'s rules align or not with \textless Platform A\textgreater{}'s rules?
    \item How do you think about \textless Platform A\textgreater{}?
    \item How does \textless Platform A\textgreater{} compare with other online community platforms?
    \item Are there other communities on \textless Platform A\textgreater{} you engage with regularly?
    \item Do you feel as though the platform rules protect you personally?
    \item Who do your community rules protect?
\end{enumerate}

\subsection*{Questions about Background Demographics}

\begin{enumerate}
    \item Do you identify as being a minority or otherwise marginalized within your culture?
    \item Do you identify as a minority or otherwise marginalized within \textless Community A\textgreater{}?
    \item How does your identity or background shape your experience with \textless Community A\textgreater{}?
    \item Do you feel like others who share your identity or background would feel comfortable participating in \textless Community A\textgreater{}? Why or why not?
    \item Would you recommend \textless Community A\textgreater{} to people with minority or otherwise marginalized identities? Why or why not?
    \item Can you describe any ways that the rules or norms of \textless Community A\textgreater{} could do more to help those with your identity or background?
    \item Can you describe any ways that the rules or norms of \textless Community A\textgreater{} already help those with your identity or background?
    \item Have you ever had a negative experience with \textless Community A\textgreater{} related to your identity or background?
\end{enumerate}

\clearpage
\newpage
\section{Rules in subreddits, Fandom wikis, and Fediverse servers}
\label{appendix:communities}
\begin{figure}[h!]
    \centering
    \includegraphics[scale=0.5]{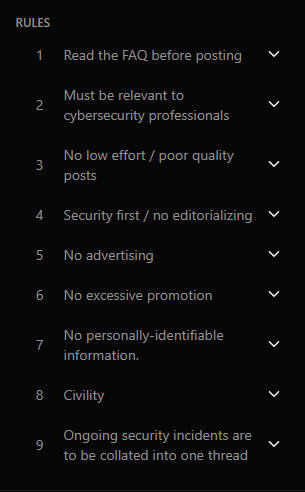}
    \caption{Rules for a subreddit community.}
    \label{fig:cybersecurityrules}
\end{figure}

\begin{figure}[h!]
    \centering
    \includegraphics[scale=0.5]{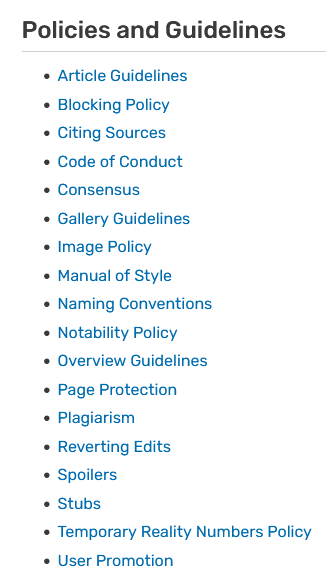}
    \caption{Rules for a Fandom wiki community.}
    \label{fig:marvelrules}
\end{figure}

\begin{figure}[h!]
    \centering
    \includegraphics[scale=0.5]{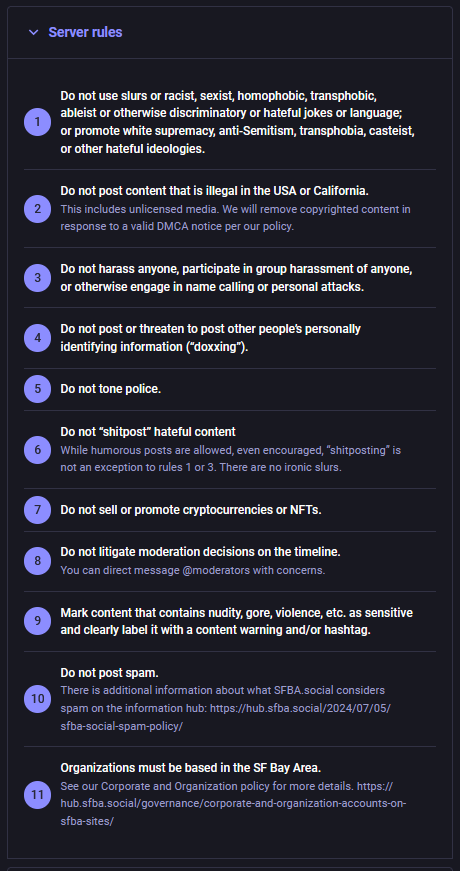}
    \caption{Rules for a Fediverse community.}
    \label{fig:sfbarules}
\end{figure}

This section provides examples of rules from online communities in each of the three settings of our study. The first set of rules is from a subreddit community (Figure \ref{fig:cybersecurityrules}). The second set of rules comes from a Fandom wiki community (Figure \ref{fig:marvelrules}). Finally, the third example of rules comes from a Fediverse community (Figure \ref{fig:sfbarules}). Communities on all three platforms tended to list additional detail about their rules, usually on other pages.

\clearpage
\newpage
\section{Participating Community Characteristics}
\label{appendix:community_tabs}
This section gives basic descriptions of the online communities from which we sampled our interview participants. We give the community's topic, when it was created, an estimate of its member population size, and the number of rules the community had adopted at the time of data collection.

\begin{table}[h!]
\centering
\begin{tabular}{lccc}
\hline
\textbf{Community Topic} & \textbf{Created} & \textbf{Size} & \textbf{Rules} \\
\hline
Television genre     & 2014     & 68889         & 11                                  \\
Q\&A     & 2011    & 20945799      & 10                                  \\
Television series         & 2008     & 202207        & 9                                   \\
IT professionals         & 2012     & 381366        & 9                                   \\
City-specific discussions     & 2008     & 60028         & 8                                   \\ Illness support         & 2013      & 41638         & 8                                   \\
Finance             & 2021     & 67428         & 15                                  \\
Book and film series           & 2008     & 1324162       & 14                                  \\
Hobbyist exchanges & 2011     & 51156         & 15                                  \\
\hline
\end{tabular} %
\caption{Reddit community characteristics. Size refers to the total number of subscribers to the subreddit.}
\label{reddit_table}
\end{table}

\begin{table}[h!]
\centering
\begin{tabular}{lccc}
\hline
\textbf{Community Topic}  & \textbf{Created} & \textbf{Size} & \textbf{Rules} \\ \hline
Fashion Styles   & 2018             & 4652          & 18                                  \\
Comic book franchise      & 2005             & 23481         & 41                                  \\
Anime fanfiction lore & 2007             & 4588          & 59                                  \\
Multiplayer video game          & 2020             & 880           & 21                                  \\
Comic book franchise           & 2009             & 15025         & 33                                  \\
Book series       & 2007             & 13365         & 135                                 \\
Book series & 2013             & 10125         & 45                                  \\
Toy franchise          & 2006             & 11926         & 28                                  \\
\hline
\end{tabular}%
\caption{Fandom community characteristics. Size refers to the total number of users who have made at least 1 contribution to the wiki.}
\label{fandom_table}
\end{table}

\begin{table}[h!]
\centering
\resizebox{\linewidth}{!}{%
\begin{tabular}{lcccc}
\hline
\textbf{Community Topic}                              & \textbf{Created} & \textbf{Size} & \textbf{Rules} & \textbf{Software} \\
\hline
City-specific and general discussions                                     & 2019             & 3700          & 7              & Mastodon \\
Region-specific and general discussions                                       & 2018             & 981           & 3              & Mastodon \\
History                          & 2018             & 26            & 4              & Mastodon \\
General discussions and news                               & 2017             & 278           & 15             & Mastodon \\
City-specific and music discussions                             & 2020             & 2             & 6              & Mastodon \\
General discussions and news                                   & Not Available    & 60           & 22        & Pleroma (Akkoma) \\
Computer hacking                                   & 2017             & 211           & 16             & Mastodon \\
Region-specific and museum discussions                                   & 2018             & 235           & 9              & Mastodon \\
Music equipment discussions                                 & 2020             & 49            & 6              & Mastodon \\
City-specific and general discussions                                   & 2018            & 370           & 11             & Mastodon \\
Musicians                                     & 2019             & 79            & 10             & Mastodon \\
Amateur hobbyists                                 & 2018             & 1100          & 9              & Mastodon \\
Video sharing                           & Not Available    &   Not Available &  Not Available & Pleroma \\
Religion-specific discussions &            Not Available         &        31          &         Not Available       & Pleroma   \\
General discussions and news                              &      Not Available               &    Not Available              &      Not Available         & Misskey    \\
Museum discussions         & 2018             & 271           & 3              & Mastodon \\
\hline
\end{tabular}%
}
\caption{Fediverse community characteristics. Size refers to the number of registered users on each server. Note: some servers' community characteristics could not be found in any public online space.}
\label{fediverse_table}
\end{table}

\end{document}